\documentstyle[prc,aps,eqsecnum,preprint,graphicx]{revtex}

\newcommand{\bra}[1]{\langle {#1} |}
\newcommand{\ket}[1]{| {#1} \rangle}
\newcommand{\inproduct}[2]{\langle #1 | #2 \rangle}

\newcommand{\Ham}{{\cal H}}
\newcommand{\Fcl}{{\cal F}}

\tighten
\begin{document}

\draft
\preprint{\parbox{4.5cm}{UMIST/Phys/TP/99-1}}
\title{Local harmonic approaches
with approximate cranking operators}
\author{Takashi~Nakatsukasa\footnote{Current address: RI Beam Science Laboratory,       RIKEN, Wako-city, Saitama 351-0198,        Japan.
       Email: ntakashi@postman.riken.go.jp}
and
Niels~R.~Walet\footnote{
Email: {Niels.Walet@umist.ac.uk}}
}
\address{Department of Physics, UMIST, P.O.Box 88,
Manchester M60 1QD, UK}
\author{Giu~Do~Dang\footnote{
Email: { Giu.Dodang@th.u-psud.fr}}}
\address{Laboratoire de Physique Th\'eorique,     
B\^at 211, Universit\'e de Paris-Sud, 91405 Orsay, France}

\maketitle

\bigskip
\begin{abstract}

Methods of large amplitude collective motion in the adiabatic limit
are examined
with a special emphasis on conservation laws.
We show that
the restriction to point transformations, which is a usual assumption
of the adiabatic time-dependent mean-field theory, needs to be lifted.
In order to facilitate the
application of large amplitude collective motion techniques,
we examine the
possibility of representing the RPA normal-mode coordinates by
linear combinations of a limited number of one-body operators.
We study the pairing-plus-quadrupole model of Baranger and Kumar
as an example, and find that such representations exist in terms of operators
that are state-dependent in a characteristic manner.
\end{abstract}


\newpage
\section{Introduction}

The selection of collective variables is an important problem
in the study of large amplitude collective motion.  In the
constrained Hartree-Fock (CHF) or Hartree-Fock-Bogoliubov (CHFB)
calculations, the collective subspaces are generated by a small number
of one-body constraint (also called generalised cranking) operators
which are most
commonly taken to be multipoles that can be represented as 
homogeneous polynomials in the coordinates, i.e.,  $r^L Y_{LK}$. In
realistic calculations of processes such as fission, the number of
coordinates needed to describe the full nuclear dynamics can easily become
larger than what can be dealt with in satisfactory manner, and a method to
determine the optimal combination needs to be devised.
Furthermore, there is no {\it a priori} reason to limit oneself to
multipole operators, and the cranking operators should be
determined by the nuclear collective dynamics itself, from amongst all
possible one-body operators.

In our past work, we have investigated a theory of adiabatic large
amplitude collective motion as a method to generate such self-consistent
collective subspaces (see Ref.~\cite{KWD91} and references
therein). The key ingredient of the method is the self-consistent
determination of the constraint operator, and as such it may provide an
answer to the selection question discussed above.
Using the local harmonic version of the theory \cite{KWD91},
the local harmonic approximation (LHA),
we have recently embarked on a study of the
properties of large amplitude collective motion in systems with
pairing.  So far we have dealt with two simple models: The first one
is a semi-microscopic model of nucleons interacting through a pairing
force, coupled to a single harmonic variable \cite{NW98_1}.  The second
\cite{NW98_2} is a fully microscopic $O(4)$ model which may be regarded as a
simplified version of the pairing-plus-quadrupole (P+Q) Hamiltonian studied
in this paper. It has turned out that the self-consistent collective
coordinate obtained by the LHA accounts quite well for the exact
dynamics of the models.  We have also shown that the CHFB calculations
using the mass-quadrupole operator as the constraint operator can
result in incorrect results \cite{NW98_2}.

In this paper, which is an  attempt to move towards realistic nuclear
problems, we will encounter some problems that need to be resolved first:
The main problem is the treatment of conservation laws.
In studies of large amplitude collective motion,
the mean-field states often do not have the symmetries of the Hamiltonian.
Then, the spurious (Nambu-Goldstone) degrees of freedom, such as
the translation and rotation of nucleus, inevitably appear,
which must be properly separated out from the other collective degrees of
freedom like nuclear shape change.
In the models discussed in Refs.~\cite{NW98_1,NW98_2},
the spurious degrees of freedom could
be removed by hand thanks to the simplicity of the models,
however in realistic nuclear problems, it is almost impossible to do so.
The second and more obvious problem
is the increase in the number of degrees of freedom.
Within the time-dependent Hartree-Bogoliubov (TDHB) approximation,
the dynamics of the models of Refs. \cite{NW98_1,NW98_2}
is described by only a modest number of degrees of freedom ($4\sim
12$). For realistic problems in heavy nuclei, on the other hand, we
need to deal with at least tens of thousands, and possibly
millions of degrees of freedom!
These two points, the effect of the conservation laws and the applications
to realistic problems,
are the main issues discussed in Sec.~\ref{sec: formalism}.

In the LHA, the collective path (or collective manifold for more than
one coordinate) is determined by solving the CHFB problem with a
cranking operator which is self-consistently determined by the local
random-phase approximation (RPA).
It is of course a well-known fact that the spurious modes decouple from the
physical modes in the RPA.
However this is true only when the system is at a minimum
of the mean-field potential.
Since the local RPA needs to be solved
at non-equilibrium points as well,
we need to study the properties of the LHA equations.
It will be shown that the symplectic version of the LHA
 combined with an extended
adiabatic approximation properly takes account of conservation laws.
It will also be shown that this formalism naturally explains the role of the
chemical potential term $-\lambda N$ in the Hamiltonian
of the quasiparticle RPA.

Since the LHA procedure requires one to solve the local RPA at large number of
 points along
the collective path (manifold), the large dimension of
the RPA matrix for realistic problems
leads to very heavy computational burdens.
Thus, instead of solving the local RPA,
we would like  to preselect a small group
of operators, and
find the RPA modes as an optimal combination of these operators at each point of the
collective surface.
This means that we restrict
the RPA diagonalisation to this small space, rather than deal with
the full millions-by-millions RPA matrix \cite{DWK94}.
In this paper, we attempt to find a set of one-body operators
which approximate well the self-consistent cranking operator.
The pairing-plus-quadrupole (P+Q) Hamiltonian is adopted for the analysis
because the RPA calculation
is relatively easy due to the separability of the force (some of the results
for the Sm isotopes have already been reported in Ref.~\cite{NWD99}).
Since the P+Q model is known to be able to describe collective
phenomena involving both pairing and quadrupole degrees of freedom
\cite{Bel59,KS60,Bes61,KS63,BK_I,BK_II,BK_III,BK_IV,BK_V},
we expect that a similar choice for the set of
one-body operators should work for other realistic Hamiltonians as well.

The P+Q model is probably one of the most simple and successful
nuclear Hamiltonians which allows us to discuss realistic problems
involving pairing and quadrupole degrees of freedom.  Baranger and
Kumar analysed in great detail the (adiabatic) collective motion in
the P+Q model assuming that the collective variables are the mass
quadrupole operators \cite{BK_I,BK_II,BK_III,BK_IV,BK_V}.
Thus, they reduced the large number
of two-quasiparticle (2qp) degrees of freedom (for this model of the order of
a thousand) to only two ``collective'' coordinates, $\beta$ and
$\gamma$.  However, our previous study of the $O(4)$ model\cite{NW98_2}
suggests that even for such simple Hamiltonians, the self-consistent
collective coordinates are often not as simple as one might think.
The P+Q model (with an additional quadrupole pairing interaction)
has also been studied by Kishimoto and Tamura \cite{KT72,KT76}
using boson expansion techniques. They found that
the coupling to (non-collective) 2qp excitations is
very important, which also suggests that the collective coordinates are
not just the mass-quadrupole ones.
In this paper we show that the
normal-mode coordinate of the random-phase approximation is
quite different from the mass-quadrupole operator,
and how the structure of the self-consistent collective coordinates 
changes as the nucleus is deformed.

In Sec.~\ref{sec: formalism}, we give an account of the general
theoretical arguments.
The problem of the spurious modes and the extension of the RPA formalism
are discussed in Sec.~\ref{sec: RLHA} and \ref{sec: SLHA}.
In Sec.~\ref{sec: PLHA}, the method of projection of the RPA equation
onto a set of one-body operators is described.
Numerical calculations for the RPA and the projected RPA for the P+Q model
are given in Sec.~\ref{sec: P_plus_Q}
and finally we give the summary and outlook in Sec.~\ref{sec: conclusions}.

\section{Local harmonic formulations and conservation laws}
\label{sec: formalism}

In this section, we use a summation convention where
the repeated appearance of the same symbol
for upper and lower index
indicates a sum over this symbol for all allowed values.
We also use the convention that a comma in a lower index
indicates the derivative with respect to the coordinate,
i.e., $F_{,\alpha} = \partial F/\partial \xi^\alpha$.

\subsection{The TDHFB theory and classical constants of motion}

Utilising the TDHFB theory
we transcribe the original quantum Hamiltonian
into the classical Hamiltonian.
We follow Ref.~\cite{BR86} to define the canonical variables.
We denote a time-dependent generalised Slater determinant by
$\ket{\Psi(t)}$ which is characterised by
the (quasi-)density and pair matrices,
\begin{equation}
\bar{\rho}_{ij} \equiv
      \bra{\Psi(t)} a_j^\dagger a_i \ket{\Psi(t)} ,\quad\quad
\bar{\kappa}_{ij} \equiv
     \bra{\Psi(t)} a_j a_i \ket{\Psi(t)} ,
\end{equation}
where $\ket{\Psi(t)}$ is assumed to be normalised at all times
($\inproduct{\Psi(t)}{\Psi(t)}=1$) and
the quasiparticle operators $(a_i, a_i^\dagger)$ are
defined with respect to a reference state $\ket{\Psi_0}$.
Based on a classical Holstein-Primakoff mapping \cite{HP40,KM90},
\begin{equation}
\label{H_P_mapping}
\bar{\rho}_{ij}
      = \left[ \beta \beta^\dagger \right]_{ij} , \quad\quad
\label{quasi_pairing_tensor}
\bar{\kappa}_{ij}
     = \left[ \beta (1 - \beta^\dagger \beta)^{1/2} \right]_{ij} ,
\end{equation}
the density and pair matrices are now mapped onto complex canonical variables
$\beta_{ij}$.
As usual real canonical variables $\xi$ and $\pi$ are introduced as
$\beta = (\xi + i \pi) / \sqrt{2} $.
In this canonical parametrisation,
the reference state $\ket{\Psi_0}$ corresponds to
the origin of the phase space, where $\xi=\pi=0$.

We shall specifically look at
one-body hermitian operators $R$ and anti-hermitian operators $S$,
\begin{eqnarray}
R &=& R_0 + \sum_{i > j} R(ij)
                          (a_i^\dagger a_j^\dagger + a_j a_i )
         + \sum_{ij} R'(ij) a_i^\dagger a_j ,\\
S &=&       \sum_{i > j} S(ij)
                          (a_i^\dagger a_j^\dagger - a_j a_i )
         + \sum_{ij} S'(ij) a_i^\dagger a_j ,
\end{eqnarray}
where $R_0=\bra{\Psi_0} R \ket{\Psi_0}$ and
all the matrix elements, $R(ij)$, $R'(ij)$, $S(ij)$ and $S'(ij)$,
are assumed to be real.
Here $R'(ij)$ ($S'(ij)$) are symmetric (anti-symmetric) with respect to
permutation of the indices $i$ and $j$.
The corresponding classical representations,
${\cal R}(\xi,\pi)$ and ${\cal S}(\xi,\pi)$, can be found by
replacing the fermion pair operators $a_j^\dagger a_i$ and $a_j a_i$
by the density and pair matrices
$\bar\rho_{ij}$ and $\bar\kappa_{ij}$ of Eq.~(\ref{H_P_mapping}).
Expanding $\bar\rho$ and $\bar\kappa$ with respect to $\pi$,
we find that ${\cal R}(\xi,\pi)$ consists of even-order terms in $\pi$
while ${\cal S}(\xi,\pi)$ contains odd-order terms only,
\begin{eqnarray}
\label{classical_R}
{\cal R} (\xi, \pi) &=&
  {\cal R}^{(0)}(\xi)
  + \frac{1}{2} {\cal R}^{(2)\alpha\beta}(\xi) \pi_\alpha \pi_\beta
  + {\cal O}(\pi^4) , \\
\label{classical_S}
{\cal S} (\xi, \pi) &=&
  {\cal S}^{(1)\alpha}(\xi) \pi_\alpha + {\cal O}(\pi^3) .
\end{eqnarray}
Starting from these equations,
pairs of two-quasiparticle (2qp) indices $(ij,kl,\cdots)$
are denoted collectively by a single Greek index ($\alpha$, $\beta$, $\cdots$).
The 2qp matrix elements with respect to the reference state,
$R(ij)$ and $S(ij)$ which are now denoted as $R(\alpha)$ and $S(\alpha)$
respectively,
are related to the derivatives of the classical representations,
\begin{eqnarray}
\label{2qp_R}
R(\alpha) &=&
    \frac{1}{\sqrt{2}}
     \left.\frac{\partial {\cal R}}{\partial \xi^\alpha}\right|_{\xi=\pi=0}
    =\frac{1}{\sqrt{2}} \left. {\cal R}^{(0)}_{,\alpha}\right|_{\xi=0} ,\\
\label{2qp_S}
S(\alpha) &=&
    \frac{i}{\sqrt{2}}
     \left.\frac{\partial {\cal S}}{\partial \pi_\alpha}\right|_{\xi=\pi=0}
    =\frac{i}{\sqrt{2}} \left. {\cal S}^{(1)\alpha}\right|_{\xi=0} .
\end{eqnarray}

The classical Hamiltonian is given by
\begin{equation}
\Ham (\xi, \pi) = \bra{\Psi(t)} H \ket{\Psi(t)} ,
\end{equation}
where the right hand side is calculated in terms of
the density and pair matrices (\ref{H_P_mapping}).
The expansion of $\Ham(\xi,\pi)$ in powers of $\pi$ defines the potential
$V(\xi) = \Ham (\pi=0)$ (zeroth order)
and the mass parameter $B^{\alpha\beta}(\xi)$ (second order).
\begin{eqnarray}
\label{adiabatic_Hamiltonian}
\Ham &=& V(\xi) + \frac{1}{2} B^{\alpha\beta} \pi_\alpha \pi_\beta
         + {\cal O}(\pi^4) ,\\
B^{\alpha\beta}(\xi) &=&
   \left. \frac{\partial^2 \Ham}{\partial\pi_\alpha \partial\pi_\beta}
    \right|_{\pi=0} .
\end{eqnarray}
The terms of order of $\pi^4$ and higher in the Hamiltonian
are neglected in the adiabatic time-dependent mean-field theory.
The tensor $B_{\alpha \beta}$, which is defined as the inverse of
$B^{\alpha \beta}$
($B^{\alpha \gamma} B_{\gamma \beta} = \delta^\alpha_\beta$),
plays the role of metric tensor in the Riemannian formulation of
the local harmonic approximation (LHA) as discussed in Sec.~\ref{sec: RLHA}.

It is well-known that the TDHFB equation preserves the symmetries
of the quantum Hamiltonian:
If the one-body operator $P$ commutes with the Hamiltonian,
$[P,H]=0$, the classical representation of $P$,
${\cal P}(\xi,\pi)=\bra{\Psi(t)} P \ket{\Psi(t)}$, is
a classical constant of motion.
Consequently, the Poisson bracket between ${\cal P}$ and $\Ham$ must vanish,
\begin{equation}
\{ {\cal P}, \Ham \}_{\rm PB} = 0 ,
\label{eq:2.12}
\end{equation}
which can be used to show that
\begin{eqnarray}
\label{TDHFB_eq_1}
{\cal P}^{(1)\alpha} V_{,\alpha} = 0 ,\\
\label{TDHFB_eq_2}
{\cal P}^{(0)}_{,\alpha} B^{\alpha\beta}
 - {\cal P}^{(2)\alpha\beta} V_{,\alpha} = 0 ,
\end{eqnarray}
which are the terms of zeroth and first order in $\pi$ in Eq.~(\ref{eq:2.12}),
respectively.
Here, ${\cal P}^{(0)}$, ${\cal P}^{(1)\alpha}$ and
${\cal P}^{(2)\alpha\beta}$ are defined,
according to Eqs.~(\ref{classical_R}) and (\ref{classical_S}),
as terms of the zeroth, first and second order in $\pi$, respectively.
The equations (\ref{TDHFB_eq_1}) and (\ref{TDHFB_eq_2})
hold at arbitrary points in configuration space.

\subsection{Riemannian version of the local harmonic approximation (LHA)}
\label{sec: RLHA}

The adiabatic approach to large amplitude collective motion is
based on a search for the collective (and non-collective) coordinates $q^\mu$
by performing a point transformation
of the original coordinates $\xi^\alpha$,
\begin{equation}
\label{point_transf_1}
q^\mu = f^\mu(\xi) , \quad\quad
\xi^\alpha = g^\alpha(q) \quad\quad (\mu,\alpha=1,\cdots,n).
\end{equation}
The conjugate momenta are given by
\begin{equation}
\label{point_transf_2}
p_\mu = g^\alpha_{,\mu} \pi_\alpha , \quad\quad
\pi_\alpha = f^\mu_{,\alpha} p_\mu .
\end{equation}
The adiabatic Hamiltonian (\ref{adiabatic_Hamiltonian}) is transformed into
\begin{eqnarray}
\bar{\Ham} &=& \bar{V}(q) + \frac{1}{2} \bar{B}^{\mu\nu} p_\mu p_\nu
         + {\cal O}(p^4) ,\\
\bar{V}(q) &=& V(g(q)) , \quad\quad
  \bar{B}^{\mu\nu} = f^\mu_{,\alpha} B^{\alpha\beta} f^\nu_{,\beta} .
\end{eqnarray}
The point transformation on the adiabatic Hamiltonian gives the same result
as the one obtained by first performing this transformation on the exact
Hamiltonian and then taking the adiabatic limit.
This is not true for a general canonical transformation,
specifically for the extended adiabatic transformation 
to be discussed in the next section.

We divide the new coordinate set $q^\mu$ into three subsets,
the collective coordinates $q^1$,
the spurious coordinates $q^I$, $I=2,\cdots,M+1$, and
the non-collective coordinates $q^a$, $a=M+2,\cdots,n$.
For simplicity, we assume here a single collective coordinate.
The spurious coordinates are related to the symmetry breaking in the
mean field (Nambu-Goldstone modes).
The local harmonic approximation  consists of two sets of equations,
the force condition and the local RPA,
which define the collective manifold in configuration space
(see Ref.~\cite{KWD91} for a complete description).
When spurious modes are present,
the force condition is modified slightly,
\begin{equation}
\label{force_condition}
V_{,\alpha} = \lambda_1 f^1_{,\alpha} + \lambda_I f^I_{,\alpha} ,
\end{equation}
where the second term in the right-hand side is the additional term.
The local RPA equation can be formulated in different ways \cite{KWD91},
and in this section we discuss the case of the Riemannian formulation,
based on
\begin{eqnarray}
\label{RLHA_1}
V_{;\alpha\gamma} B^{\gamma\beta} f^1_{,\beta}
   &=& (\hbar\Omega)^2 f^1_{,\alpha}, \\
\label{RLHA_2}
g^\alpha_{,1} V_{;\alpha\gamma} B^{\gamma\beta}
   &=& (\hbar\Omega)^2 g^\beta_{,1}.
\end{eqnarray}
Here the covariant derivative $V_{;\alpha\gamma}$ is defined by
\begin{equation}
V_{;\alpha\beta} \equiv V_{,\alpha\beta}
               - \Gamma^\gamma_{\alpha\beta} V_{,\gamma} ,
\end{equation}
where the affine connection $\Gamma$ is defined with the help of
metric tensor $B_{\alpha\beta}$ as
\begin{equation}
\label{affine_1}
\Gamma^\alpha_{\beta\gamma} = \frac{1}{2}
   B^{\alpha\delta} \left( B_{\delta\beta, \gamma}
                        + B_{\delta\gamma , \beta}
                        - B_{\beta\gamma , \delta} \right) .
\end{equation}
The collective path will be
determined by solving Eqs.~(\ref{force_condition}) and (\ref{RLHA_1})
selfconsistently.
It consists in finding a series of points where the local RPA eigenvector
$f^1_{,\alpha}$ satisfies the force condition at the same time.

The question at present is whether the solution of Eqs.~(\ref{RLHA_1}) and
(\ref{RLHA_2}) is orthogonal to the spurious modes.
The spurious coordinates are normally linked to the one-body
operators $P_I$ which commute with the Hamiltonian.
It is convenient to discuss separately the case
in which the symmetry operators $P_I$
are hermitian with real matrix elements
and the case in which $P_I$ are anti-hermitian with
real matrix elements.
First, let us discuss the latter case.
As is seen in Eq.~(\ref{classical_S}), neglecting the higher-order terms
in $\pi$,
the classical representation of the symmetry operator
is a classical momentum variable $p_I$ that is a constant of motion,
\begin{equation}
p_I =  \bra{\Psi} P_I \ket{\Psi}
   = {\cal P}_I^{(1)\alpha} \pi_\alpha + {\cal O}(\pi^3) .
\end{equation}
From this, we can immediately see that
$g^\alpha_{,I} = {\cal P}_I^{(1)\alpha}$.
Differentiating Eq.~(\ref{TDHFB_eq_1}) with respect to $\xi^\beta$,
keeping in mind that
${\cal P}^{(1)\alpha}$ must be proportional to $g^\alpha_{,I}$,
we find
\begin{equation}
\label{TDHFB_eq_3}
V_{,\alpha\beta} g^\alpha_{,I}
 + V_{,\alpha} g^\alpha_{,I\mu} f^\mu_{,\beta} = 0 .
\end{equation}
Using an identity for the coordinate transformation \cite{KWD91},
\begin{equation}
g^\alpha_{,\mu\nu}
            + \Gamma^\alpha_{\beta\gamma} g^\beta_{,\mu} g^\gamma_{,\nu}
            - \bar{\Gamma}^\lambda_{\mu\nu} g^\alpha_{,\lambda} = 0 ,
\end{equation}
this leads to
\begin{equation}
\label{RLHA_NG_1}
g^\alpha_{,I} V_{;\alpha\gamma} B^{\gamma\beta} =
 - \bar{V}_{,\lambda} \bar{\Gamma}^\lambda_{I\mu}
                  \bar{B}^{\mu\nu} g^\beta_{,\nu} ,
\end{equation}
where $\bar{\Gamma}$ is defined in the same way as $\Gamma$,
(\ref{affine_1}), with obvious change of $B$ to $\bar{B}$.
At the equilibrium $V_{,\alpha} = \bar{V}_{,\mu} = 0$,
and since the right-hand side of Eq.~(\ref{RLHA_NG_1}) vanishes,
the spurious modes are indeed
the zero-energy solutions of the RPA equation.
Thus, all finite-energy solutions are automatically
orthogonal to the spurious modes.
However, at non-equilibrium points, the $g^\alpha_{,I}$ are no longer
RPA eigenmodes and there is no guarantee that the collective mode
$g^\alpha_{,1}$ obtained from the local RPA equation (\ref{RLHA_2})
is free from spurious admixture\footnote{
If  $\bar{B}^{\mu\nu}$ and $\bar{\Gamma}^1_{\mu\nu}$
are block-diagonal for the spurious and the collective spaces
on the collective path,
the spurious solutions correspond to finite energy eigenvalues but
are orthogonal to the collective mode.
This is the case for HF problem in $^{28}$Si as discussed in Refs.~\cite{DWK94,WDK91,WKD92}.}.
In Ref.~\cite{NW98_1}, the local RPA equation was modified to
insure that $f^I_{,\alpha} g^\alpha_{,1} = g^\alpha_{,I} f^1_{,\alpha} =0$.
However, in this formulation, we need to calculate both $f^I_{,\alpha}$
and $g^\alpha_{,I}$, which means that we have to solve the
Thouless-Valatin equations for the spurious modes \cite{TV62,MW69}.

In the second case, where the symmetry operator $P_I$ is hermitian,
we have coordinates $q^I$ which are the constants of motion:
\begin{equation}
\label{q_point_transf}
q^I =  \bra{\Psi} P_I \ket{\Psi}
   = {\cal P}_I^{(0)}(\xi)
   + \frac{1}{2} {\cal P}_I^{(2)\alpha\beta} \pi_\alpha \pi_\beta
   + {\cal O}(\pi^4) .
\end{equation}
If we limit ourselves to point transformations, we find
${\cal P}_I^{(0)}(\xi) = f^I (\xi)$ and  we should ignore all terms depending on $\pi$.
At equilibrium Eq.~(\ref{TDHFB_eq_2}) implies that
the spurious modes $f^I_{,\alpha}$ correspond to
zero eigenmode of the mass matrix $B^{\alpha\beta} f^I_{,\beta} = 0$.
Therefore we have a zero-energy mode,
\begin{equation}
\label{RLHA_NG_2}
   V_{;\beta\gamma} B^{\gamma\alpha} f^I_{,\alpha} = 0 .
\end{equation}
Away from equilibrium, however, the $f^I_{,\alpha}$ are no longer RPA eigenmodes, due to the second term in (\ref{TDHFB_eq_2})
which is related to the terms we have neglected in Eq.~(\ref{q_point_transf}).

In summary, the Riemannian LHA can automatically separate out the spurious
modes only at equilibrium $V_{,\alpha}=0$.
Since the local RPA at equilibrium is nothing but the conventional RPA,
the decoupling of the spurious modes is well-known.
However, at non-equilibrium points, the spurious modes can in general mix
with the collective one.
This undesirable mixing disappears in the symplectic LHA with an extended
adiabatic treatment described in the next section.

\subsection{Symplectic LHA and extended adiabatic transformation}
\label{sec: SLHA}

The symplectic version of the LHA \cite{KWD91}
is formulated with an affine connection
\begin{equation}
\widetilde{\Gamma}^\alpha_{\beta\gamma}
 \equiv g^\alpha_{,\mu} f^\mu_{,\beta\gamma} ,
\end{equation}
which can be written in the usual form (\ref{affine_1}) by replacing
the metric tensor $B_{\alpha\beta}$ by
\begin{equation}
K_{\alpha\beta} \equiv \sum_\mu f^\mu_{,\alpha} f^\mu_{,\beta} .
\end{equation}
Since this metric depends on the final coordinates $q^\mu$,
we cannot define it from the beginning in contrast to the
Riemannian metric $B_{\alpha\beta}$.
In this formulation,
the configuration space is
assumed to be flat and diagonal ($K_{\mu\nu} = \delta_{\mu\nu}$)
in the final coordinate system $q^\mu$.
The covariant derivative is defined by
\begin{eqnarray}
\label{SLHA_cov_dev}
\widetilde{V}_{;\alpha\beta} &=&
  V_{,\alpha\beta}
   - \widetilde{\Gamma}^\gamma_{\alpha\beta} V_{,\gamma} \nonumber \\
 &=& V_{,\alpha\beta} - f^\mu_{,\alpha\beta} \bar{V}_{,\mu} .
\end{eqnarray}

For the case that the momenta $p_I$ are constants of motion,
we can prove that the spurious modes are the zero-energy solutions
of the symplectic local RPA equation.
Indeed, differentiating the chain relation
$g^\alpha_{,\mu} f^\mu_{,\beta} = \delta^\alpha_\beta$
with respect to $q^\nu$, we obtain
\begin{equation}
g^\alpha_{,\mu\nu} f^\mu_{,\beta}
  = - g^\alpha_{,\mu} g^\gamma_{,\nu} f^\mu_{,\beta\gamma} .
\end{equation}
Utilising this equation and the definition of the covariant derivatives
(\ref{SLHA_cov_dev}),
Eq.~(\ref{TDHFB_eq_3}) can be rewritten as
\begin{equation}
\widetilde{V}_{;\alpha\beta} g^\alpha_{,I} = 0 .
\end{equation}
This is a consequence of the TDHFB equation of motion.
Therefore the spurious modes are zero-energy solutions of the
symplectic LHA equations, anywhere in the configuration space.

In the case that the coordinates $q^I$ are constants of motion,
we need to lift the restriction to point transformations,
based on the power expansion with respect to $\pi$ keeping the
adiabatic assumption.
Instead of Eqs.~(\ref{point_transf_1}) and (\ref{point_transf_2}),
we generalise these to
\begin{eqnarray}
\label{EAT_0}
q^\mu &=& f^\mu(\xi)
  + \frac{1}{2} f^{(1)\mu\alpha\beta}(\xi) \pi_\alpha \pi_\beta
  + {\cal O}(\pi^4) , \\
\label{EAT_0b}
\xi^\alpha &=& g^\alpha(q)
  + \frac{1}{2} g^{(1)\alpha\mu\nu}(q) p_\mu p_\nu
  + {\cal O}(p^4) ,
\end{eqnarray}
and
\begin{eqnarray}
p_\mu &=& g^\alpha_{,\mu} \pi_\alpha + {\cal O}(\pi^3) ,\\
\pi_\alpha &=& f^\mu_{,\alpha} p_\mu + {\cal O}(p^3) ,
\label{eq:2.37}
\end{eqnarray}
where the terms cubic in momenta do not play a role in the modification
of the theory.
The original Hamiltonian (\ref{adiabatic_Hamiltonian}) is transformed to,
up to second order in $p$ only,
\begin{eqnarray}
\bar{H}(q,p) &=& \bar{V}(q) + \frac{1}{2} \bar{B}^{\mu\nu} p_\mu p_\nu ,\\
\bar{B}^{\mu\nu} &=& f^\mu_{,\alpha} B^{\alpha\beta} f^\nu_{,\beta}
                     + V_{,\gamma} g^{(1)\gamma\mu\nu} .
\end{eqnarray}
Substituting Eq.~(\ref{EAT_0}) in Eq.~(\ref{EAT_0b}), using Eq.~(\ref{eq:2.37}),
we find the relation
\begin{equation}
\label{EAT_1}
g^{(1)\alpha\mu\nu} g^\beta_{,\mu} g^\gamma_{,\nu} =
 -f^{(1)\lambda\beta\gamma} g^\alpha_{,\lambda} .
\end{equation}
From the canonicity condition $\{ q^\mu, q^\nu \}_{\rm PB} = 0$,
we also find
\begin{equation}
\label{EAT_2}
f^\mu_{,\alpha} f^{(1)\nu\alpha\beta}
 = f^\nu_{,\alpha} f^{(1)\mu\alpha\beta} .
\end{equation}

The major difference between the use of the extended adiabatic
transformation and a point transformation is the modification
of mass parameter,
\begin{eqnarray}
\label{modified_mass}
\widetilde{B}^{\alpha\beta} &\equiv&
   g^\alpha_{,\mu} \bar{B}^{\mu\nu} g^\beta_{,\nu} \nonumber \\
 &=& B^{\alpha\beta} - \bar{V}_{,\mu} f^{(1)\mu\alpha\beta} .
\end{eqnarray}
Here we have used the relation (\ref{EAT_1}) to obtain the last equation.
The local RPA equations have the same forms as the Riemannian RPA equations,
(\ref{RLHA_1}) and (\ref{RLHA_2}),
after replacing $V_{;\alpha\beta}$ by $\widetilde{V}_{;\alpha\beta}$ and
$B^{\alpha\beta}$ by $\widetilde{B}^{\alpha\beta}$.

Let us return to the spurious modes.
For hermitian constants of motion, i.e., coordinates $q^I$ 
as in Eq.~(\ref{q_point_transf}),
we can identify $f^I(\xi) = {\cal P}^{(0)}_I(\xi)$ and
$f^{(1)I\alpha\beta} = {\cal P}^{(2)\alpha\beta}_I$
for the extended adiabatic transformation.
Utilising the TDHFB equation (\ref{TDHFB_eq_2})
and the canonicity condition (\ref{EAT_2}), we find
\begin{eqnarray}
\label{EAT_3}
\widetilde{B}^{\alpha\beta} f^I_{,\beta} &=&
  B^{\alpha\beta} f^I_{,\beta}
   - f^{(1)\mu\alpha\beta} \bar{V}_{,\mu} f^I_{,\beta} \nonumber \\
 &=& B^{\alpha\beta} f^I_{,\beta}
   - f^{(1)I\alpha\beta} \bar{V}_{,\mu} f^\mu_{,\beta} \nonumber \\
 &=& B^{\alpha\beta} f^I_{,\beta}
   - f^{(1)I\alpha\beta} V_{,\beta} \nonumber \\
 &=& 0 .
\end{eqnarray}
Therefore, anywhere in the space, it is guaranteed that
the spurious modes are zero-energy solutions of
the symplectic LHA with the extended adiabatic transformation.

It is illustrative to discuss the difference between
the Hartree-Fock (HF) and Hartree-Fock-Bogoliubov (HFB) approaches.
The HF state is at an equilibrium point of the potential, $V_{,\alpha}=0$,
where the force condition (\ref{force_condition}) is satisfied
with $\lambda_1=\lambda_I=0$.
We also see
$\widetilde{V}_{;\alpha\beta} = V_{;\alpha\beta} = V_{,\alpha\beta}$
and $\widetilde{B}^{\alpha\beta} = B^{\alpha\beta}$.
Thus, any version of the LHA is equivalent to the conventional RPA.
On the other hand, the HFB state is not a real equilibrium.
We have a spurious coordinate associated with the particle number
$q^N = \bra{\Psi} N \ket{\Psi}$
and the gradient of the potential has a non-zero component
along the direction of $q^N$, $\bar{V}_{,N} \neq 0$
(the other components of the gradient are all zero).
The force condition is again satisfied with $\lambda_1=0$ and
$\lambda_N=\bar{V}_{,N}$, but neither
$\widetilde{V}_{;\alpha\beta} = V_{;\alpha\beta}$ nor
$\widetilde{B}^{\alpha\beta} = B^{\alpha\beta}$ hold any more.
Instead, we have
\begin{eqnarray}
\label{EAT_HFB}
\widetilde{B}^{\alpha\beta} &=& B^{\alpha\beta}
   - \bar{V}_{,N} f^{(1)N\alpha\beta} ,\\
\label{SLHA_HFB}
\widetilde{V}_{;\alpha\beta} &=& V_{,\alpha\beta}
   - \bar{V}_{,N} f^N_{,\alpha\beta} .
\end{eqnarray}
Since $\bar{V}_{,N}$ is nothing but the chemical potential $\lambda_N$
at the (constrained) HFB minimum,
the symplectic version of the LHA with the extended adiabatic approximation
is equivalent to the conventional quasiparticle RPA
with the constrained Hamiltonian, $H' = H - \lambda_N N$.
This guarantees the separation of zero-energy spurious modes,
while this is not the case for the Riemannian version of the LHA.
This symplectic formulation also illuminates
why the quasiparticle RPA at the HFB state
should be performed using the Hamiltonian $H'$, not the original $H$.

As we have seen above, the symplectic formulation of LHA with the extended
adiabatic transformation has an advantage over its Riemannian version.
Unfortunately we do not know a general method to calculate
$f^{(1)\mu\alpha\beta}$.
It may be possible to calculate this quantity
if the collective coordinate is explicitly given by 
a combination of elementary one-body operators (see the discussion in Sec.~\ref{sec: conclusions}).

\subsection{Projected LHA and the spurious modes}
\label{sec: PLHA}

One of the main obstacles in applying the LHA techniques described in previous
sections to realistic nuclear problems is the fact that we need to diagonalise
an RPA matrix of large dimensionality at each point of the collective path.
In this section, we describe a method to approximate the RPA eigenvector
by taking linear combinations of preselected one-body operators,
generalising the method discussed in Ref. \cite{DWK94}.

First we select a small number of one-body operators,
$F^{(k)}$, $k=1,\cdots,n'$,
assuming that the collective coordinate
$q^1$
can be approximated locally as a
linear combination of these operators,
\begin{equation}
q^1 \approx \bra{\Psi} \hat{f} \ket{\Psi} , \quad
\hat{f} = \sum_{k=1}^{n'} c_k F^{(k)} .
\end{equation}
This means that the RPA eigenmodes $f_{,\alpha}$ is projected onto
the subspace spanned by $\{ \Fcl^{(k)}_{,\alpha} \}$,
\begin{equation}
f_{,\alpha} \approx \bar{f}_{,\alpha} = \sum_{k=1}^{n'} c_k \Fcl^{(k)}_{,\alpha} ,
\end{equation}
where $\Fcl^{(k)}$ are the classical representations
(expectation values) of $F^{(k)}$.
In order to determine the coefficients $c_k$ in the linear combinations,
the local RPA equation is also projected onto the subspace
$\{ \Fcl^{(k)}_{,\alpha} \}$:
\begin{equation}
\label{projected_LHA}
{\bf M}^{kl} c_l = (\hbar\bar{\Omega})^2 {\bf N}^{kl} c_l ,\\
\end{equation}
where $\hbar\bar{\Omega}$ is a frequency of the projected RPA and
\begin{eqnarray}
\label{projected_LHA_M}
{\bf M}^{kl} &=& \Fcl^{(k)}_{,\alpha} B^{\alpha\beta}
             V_{;\beta\gamma} B^{\gamma\delta} \Fcl^{(l)}_{,\delta} ,\\
\label{projected_LHA_N}
{\bf N}^{kl} &=& \Fcl^{(k)}_{,\alpha} B^{\alpha\beta} \Fcl^{(l)}_{,\beta} .
\end{eqnarray}
Normally, in order to obtain the RPA eigenmodes and frequencies,
we need to diagonalise the RPA matrix
$B^{\alpha\gamma} V_{;\gamma\beta}$
whose dimension is equal to the number of active 2qp degrees of freedom.
The dimension of matrices ${\bf M}^{kl}$ and
${\bf N}^{kl}$ is equal to the number of selected one-body
operators $\{F^{(k)}\}$.
Therefore, if we can approximate the RPA
eigenvectors by using a small number of operators, it will
significantly reduce the computational task.

The equation for the force condition is also projected the same way,
\begin{equation}
\label{projected_force_condition_1}
{\bf V}^{k} = \lambda_1 {\bf N}^{kl} c_l ,\\
\end{equation}
with
\begin{equation}
\label{projected_LHA_V}
{\bf V}^{k} = \Fcl^{(k)}_{,\alpha} B^{\alpha\beta} V_{,\beta} .
\end{equation}
Equations (\ref{projected_LHA}) and
(\ref{projected_force_condition_1}) must be solved self-consistently
in order to determine a collective path.

In cases where spurious modes exist,
the spurious components should be removed from each operator as
\begin{equation}
\label{PLHA_op}
\bar\Fcl^{(k)}_{,\alpha} = \Fcl^{(k)}_{,\alpha}
     - \Fcl^{(k)}_{,\beta} g^\beta_{,I} f^I_{,\alpha} ,
\end{equation}
and the force condition (\ref{projected_force_condition_1}) should be also
modified by adding the terms $\lambda_I {\bf W}^{kI}$ in the right-hand
side with
${\bf W}^{kI} = \bar\Fcl^{(k)}_{,\alpha} B^{\alpha\beta} f^I_{,\beta}$.

The removal of spurious modes in Eq.~(\ref{PLHA_op}) is a tedious task
because we need to obtain both $f^I_{,\alpha}$ and $g^\alpha_{,I}$
by solving the Thouless-Valatin equation.
However, in case of hermitian constants of motion $q^I$,
adopting the extended adiabatic approximation in the previous section,
we show that the explicit removal as in Eq.~(\ref{PLHA_op}) is unnecessary.
Using the modified mass parameter (\ref{modified_mass}),
we have $\widetilde{B}^{\alpha\beta} f^I_{,\beta}=0$, Eq.~(\ref{EAT_3})
showing that 
the $f^I_{,\alpha}$ correspond to the zero eigenmodes of mass matrix
(\ref{EAT_3}).
Therefore, we find
\begin{eqnarray}
{\bf M}^{kl} &=& \Fcl^{(k)}_{,\alpha} B^{\alpha\beta}
            V_{;\beta\gamma} B^{\gamma\delta} \Fcl^{(l)}_{,\delta} 
            = \bar\Fcl^{(k)}_{,\alpha} B^{\alpha\beta}
            V_{;\beta\gamma} B^{\gamma\delta} \bar\Fcl^{(l)}_{,\delta},\\
{\bf N}^{kl} &=& \Fcl^{(k)}_{,\alpha} B^{\alpha\beta} \Fcl^{(l)}_{,\beta} 
            = \bar\Fcl^{(k)}_{,\alpha} B^{\alpha\beta} \bar\Fcl^{(l)}_{,\beta}
            ,\\
{\bf V}^k    &=& \Fcl^{(k)}_{,\alpha} B^{\alpha\beta} V_{,\beta}
            = \bar\Fcl^{(k)}_{,\alpha} B^{\alpha\beta} V_{,\beta} ,
\end{eqnarray}
and ${\bf W}^{kI}= 0$,
which means that
the spurious components of one-body operators $F^{(k)}$
do not play any role  in the projected LHA formalism.
All we need to do is to keep $q^I$ constant all the way
along the collective path.
Then, we can use the original set of one-body operators $\{ F^{(k)} \}$
on which the LHA equations are projected, without modifying
Eqs.~(\ref{projected_LHA}) to (\ref{projected_LHA_V}).

In the classical theory of the P+Q model discussed in the next section
the constants of motion are coordinates $q^I$, and not momenta $p_I$.
Therefore this is exactly the case where the mixing of spurious components
does not play any role.
For further details, see the discussion around Eq.~(\ref{PQ_spurious}).

\section{Collective coordinates for the P+Q model}
\label{sec: P_plus_Q}

In this section,
we investigate the structure of the self-consistent collective coordinates
(cranking operators) at the Hartree-Bogoliubov (HB) state
for the pairing-plus-quadrupole (P+Q) Hamiltonian,
as a first step towards the large amplitude collective motion
in heavy nuclei.
We also try to find a set of one-body operators which can approximate the
RPA eigenvectors.

\subsection{The quasiparticle RPA for the P+Q model}

We apply the symplectic version of LHA with the extended adiabatic
approximation to the P+Q Hamiltonian.
At the HB state, as shown in Sec.~\ref{sec: SLHA},
the force condition can be always satisfied and the local RPA equation
is equivalent to the quasiparticle RPA for the constrained Hamiltonian
\begin{eqnarray}
H' &=& H - \sum_{\tau=n,p} \lambda_\tau N_\tau ,\\
H  &=& \sum_k \epsilon_k c_k^\dagger c_k
      -\sum_{\tau=n,p} \frac{G_\tau}{2}
           \left( P_\tau^\dagger P_\tau + P_\tau P_\tau^\dagger \right)
      -\frac{\chi}{2} \sum_{K=-2}^2 Q_{2K}^\dagger Q_{2K} \\
   &=& \sum_k \epsilon_k c_k^\dagger c_k
      -\frac{1}{2} \sum_\sigma \kappa_\sigma R_\sigma R_\sigma
      +\frac{1}{2} \sum_\sigma \kappa_\sigma S_\sigma S_\sigma ,
\end{eqnarray}
where $\epsilon_k$ are spherical single-particle energies and
$N_\tau = \sum_{k\in\tau} c_k^\dagger c_k$ are the number operators
for neutrons ($\tau=n$) and protons ($\tau=p$).
The operators
$R_\sigma$ and $S_\sigma$ are the hermitian and anti-hermitian
components, respectively, of the pairing operators,
$P_\tau^\dagger=\sum_{k\in\tau, k>0} c_k^\dagger c_{\bar k}^\dagger$,
and the dimensionless quadrupole operators,
$Q_{2K} = b_0^{-2} \sum_{kl} \bra{k} r^2 Y_{2K} \ket{l} c_k^\dagger c_l$,
where $b_0=(\hbar/m\omega_0)^{1/2}$ is the harmonic oscillator length.
The Hamiltonian contains five operators of $R_\sigma$-type and four of
$S_\sigma$-type.
Together with the corresponding coupling constants $\kappa_\sigma$,
they are given by
\begin{equation}
\label{PQ_table}
\begin{array}{ccccccc}
R_\sigma &=& (P_+^{(+)})_n, & (P_+^{(+)})_p,
              &Q^{(+)}_{20}, &Q^{(-)}_{21}, &Q^{(+)}_{22},\\
S_\sigma &=& (P_-^{(+)})_n, & (P_-^{(+)})_p,& &Q^{(+)}_{21}, &Q^{(-)}_{22},\\
\kappa_\sigma &=& G_n, & G_p, &\chi, &\chi, &\chi,
\end{array}
\end{equation}
where
\begin{eqnarray}
(P_\pm^{(+)})_\tau &=& \frac{1}{\sqrt{2}} ( P_\tau \pm P_\tau^\dagger ),
           \quad \mbox{ for } \tau=n,p, \nonumber \\
\label{sig_good_Q}
Q^{(\pm)}_{2K} &=&  \frac{1}{\sqrt{2}} ( Q_{2K} \pm Q_{2-K} ) , \quad
\mbox{ for } K=0,1,2.
\end{eqnarray}
The signs $(\pm)$ indicate the signature quantum number,
$e^{-i\pi J_x} O^{(\pm)} e^{i\pi J_x} = \pm O^{(\pm)}$.
Following the standard formulation of the model,
we will neglect the Fock terms,
the contributions of the pairing force to the Hartree potential
and those of the quadrupole force to the pairing potential
\cite{Bel59,KS60,Bes61,KS63,BK_I,BK_II,BK_III,BK_IV,BK_V}.
After minimising the HB total energy
and diagonalising the HB matrix,
we obtain
the HB ground state $\ket{\Psi_0}$ and the quasiparticle energies $E_k$.
Using $\ket{\Psi_0}$ as the reference state,
the classical Hamiltonian is calculated as
\begin{eqnarray}
\label{H'_PQ}
\Ham'(\xi,\pi) &=& \Ham(\xi,\pi) 
         - \sum_\tau \lambda_\tau {\cal N}_\tau (\xi,\pi) \nonumber\\
  &=& E_0' + \sum_k E_k \bar\rho_{kk}
   - \frac{1}{2} \sum_\sigma \kappa_\sigma
     \left( \sum_{i>j} R_\sigma(ij) (\bar\kappa_{ij}^* + \bar\kappa_{ij})
         +\sum_{ij} R'_\sigma(ij) \bar\rho_{ji} \right)^2 \nonumber \\
  &&\hspace*{2.5cm} +\frac{1}{2} \sum_\sigma \kappa_\sigma
  \left( \sum_{i>j} S_\sigma(ij) (\bar\kappa_{ij}^* - \bar\kappa_{ij})
         +\sum_{ij} S'_\sigma(ij) \bar\rho_{ji} \right)^2 .
\end{eqnarray}
Applying the Holstein-Primakoff mapping (\ref{H_P_mapping})
and taking into account that the $R'(ij)$ are symmetric and
$S'(ij)$ are anti-symmetric with respect to the indices $(ij)$,
one can see that $\bra{\Psi(t)} R_\sigma \ket{\Psi(t)}$
($\bra{\Psi(t)} S_\sigma \ket{\Psi(t)}$) are real (pure imaginary),
and indeed the Hamiltonian (\ref{H'_PQ}) is real.
Differentiating $\Ham'$ with respect to $\pi$ and $\xi$,
the mass and curvature parameters at the HB state, defined by
Eqs.~(\ref{EAT_HFB}) and (\ref{SLHA_HFB}), are given by
\begin{eqnarray}
\label{B_PQ}
\widetilde{B}^{\alpha\beta}(\xi=\pi=0) &=&
 B^{\alpha\beta} - \sum_{\tau=n,p} \lambda_\tau f^{(1)N_\tau \alpha\beta}
 \nonumber \\
 &=& E_\alpha \delta_{\alpha\beta}
   - 2\sum_\sigma \kappa_\sigma S_\sigma(\alpha) S_\sigma(\beta) ,\\
\label{V_PQ}
\widetilde{V}_{;\alpha\beta}(\xi=\pi=0) &=&
 V_{,\alpha\beta} - \sum_{\tau=n,p} \lambda_\tau f^{N_\tau}_{,\alpha\beta}
 \nonumber \\
 &=& E_\alpha \delta_{\alpha\beta}
   - 2\sum_\sigma \kappa_\sigma R_\sigma(\alpha) R_\sigma(\beta) .
\end{eqnarray}
Here, again the Greek index indicates a pair of 2qp index,
so that $R(\alpha)$ are the 2qp matrix elements $R(\alpha)= R(ij)$ and
$E_\alpha$ are the 2qp energies $E_\alpha = E_i+E_j$.
We see that the terms $R'(ij)\bar{\rho}_{ji}$ and $S'(ij)\bar{\rho}_{ji}$
in the Hamiltonian do not contribute to
the mass and curvature parameters, which is consistent with the fact
that the terms proportional to $(a^\dagger a)$ in one-body operators
are neglected in the quasiparticle RPA.

Solving the RPA is equivalent
to diagonalisation of the matrix
$\widetilde{B}^{\alpha\gamma} \widetilde{V}_{;\gamma\beta}$.
For separable forces such as the P+Q model,
the problem of diagonalisation can be reduced to a root search
in a multi-dimensional dispersion equation,
which facilitates the numerical calculations for heavy nuclei \cite{RS80}.
If the ground state has an axial symmetry, the RPA matrix is block-diagonal
and can be divided according to $K$ quantum numbers (eigenvalues of $J_z$).
However, as we will show below, some nuclei have triaxial HB ground states
for which $K$ is no longer a good quantum number.
Thus, we use the signature quantum number $r$,
which is good even for triaxial cases, and
the RPA dispersion equation can be divided
into positive ($r=+1$) and negative ($r=-1$) signature sectors.
The construction of the response functions
and the coupled dispersion equations
can be found, for example, in Ref. \cite{NMMS96}.

\subsection{Numerical results}
\label{sec: results}

We follow the second and third of the series of papers by Baranger and Kumar
\cite{BK_II,BK_III} for details of the P+Q model.
The model space consists of two major shells of $N_{\rm osc} = 5, 6$
and $4, 5$ for neutrons and protons, respectively.
Different harmonic oscillator frequencies are used for neutrons
and protons in order to make the root mean square radii the same \cite{BK_II}.

The single-particle energies
are taken from Table~1 in Ref. \cite{BK_III},
where they are listed in units of the
harmonic oscillator frequency $\hbar\omega_0=41.2A^{-1/3}$ MeV.
The pairing force strengths are
$G_n=22A^{-1}$ MeV and $G_p=27A^{-1}$ MeV
for neutrons and protons, respectively.
The quadrupole force strength is $\chi=70 A^{-1.4}$ MeV.
The effective charge in the $E2$ operator is taken as $e_n = 1.5 Z A^{-1}$
and $e_p = 1 + e_n$ for neutrons and protons, respectively.
Since matrix elements of the quadrupole force in upper shells are known
to be too strong \cite{BK_II},
we modify the quadrupole operators
by multiplying all quadrupole matrix
elements by a factor
$\zeta = (N_{\rm L}+\frac{3}{2})/(N_{\rm osc}+\frac{3}{2})$,
where $N_{\rm osc}$ is the oscillator quantum number of the shell under
consideration and $N_{\rm L}$ is that of the lower major shell.
Furthermore,
due to the difference of harmonic oscillator frequencies
for neutrons and protons,
the operators are multiplied by factors $\alpha_n^2=(2N/A)^{2/3}$
and $\alpha_p^2=(2Z/A)^{2/3}$ for neutrons and protons, respectively.
Thus,
the dimensionless quadrupole operators in the interaction (\ref{PQ_table})
are actually defined by
\begin{eqnarray}
\label{modify_Q}
Q_{2K} &=& (Q_{2K})_n + (Q_{2K})_p ,\nonumber\\
(Q_{2K})_n &=& \alpha_n^2 \zeta b_0^{-2} (r^2 Y_{2K})_n , \quad
(Q_{2K})_p  =  \alpha_p^2 \zeta b_0^{-2} (r^2 Y_{2K})_p .
\end{eqnarray}
These operators will be referred to simply as the ``quadrupole operators''
in the following discussion.

We have performed calculations for even-even nuclei
in the rare-earth region ($N=82\sim 126$, $Z=50\sim 82$)
and have investigated the structure of RPA eigenvectors.
Since $\beta$ and $\gamma$ vibrations in these nuclei were studied by
Baranger and Kumar in Ref. \cite{BK_III},
we can compare our microscopic RPA results
with their semi-microscopic ones.

In Table~\ref{HFB_plus_RPA}, results of the HB and the RPA
calculations are presented.
The equilibrium deformations ($\beta$, $\gamma$, $\Delta_n$, $\Delta_p$)
at HB ground states are found to agree with Table~2 in Ref. \cite{BK_III},
except $^{190}$Pt which has a small triaxiality ($\gamma=52.6^\circ$)
in our result but has a oblate shape ($\gamma=60^\circ$) in theirs.
The deformation parameters, $\beta$ and $\gamma$, of the HB states
are defined by \cite{BK_II,BK_III},
\begin{equation}
 \chi \bra{\Psi_0} Q_{20}^{(+)} \ket{\Psi_0}
   = \hbar\omega_0 \beta \cos\gamma ,\quad
 \chi \bra{\Psi_0} Q_{22}^{(+)} \ket{\Psi_0}
   = \hbar\omega_0 \beta \sin\gamma .
\end{equation}
We have solved the RPA in the positive signature sector where
the spurious modes, the pairing and spatial rotation, are coordinates
(not momenta) and are characterised
by zero eigenmodes of the mass matrix,
$B^{\alpha\beta} f^I_{,\beta} = 0$.
For spherical nuclei ($\beta=0$), the lowest excited states
with multipolarity $\lambda=2$ are selected
as quadrupole vibrations and presented in columns 7 and 8.
For prolate ($\beta>0,\gamma=0$) and oblate ($\beta<0,\gamma=0$) deformed
cases, the first excited states (except zero spurious modes)
with $K=0$ and $K=2$ are
taken as $\beta$ and $\gamma$ vibrations, respectively.
However, for some cases, the lowest excited states appear to be
non-collective and we have selected higher-energy states.
These states are indicated with $*$ in the table.
For a triaxial ground state, the first excited states are selected and
the excitation energies are shown in column 7 while the transition
amplitudes are given in columns 8 and 10.
The $E2$ amplitudes $M(E2)_K$ are intrinsic values and calculated as
\begin{equation}
M(E2)_K =
 \left| \bra{n^{(+)}} e_n \zeta (r^2Y_{2K})^{(+)}_n
                     + e_p \zeta (r^2Y_{2K})^{(+)}_p \ket{0} \right| ,
\end{equation}
where $\ket{0}$ and $\ket{n^{(+)}}$ are the RPA ground and excited states
with signature $r=+1$, respectively.
For spherical nuclei, states with $K=0$, 1 and 2
are degenerate in energy and we select the $K=0$ one for $\ket{n^{(+)}}$.
For triaxial shapes, the state $\ket{n^{(+)}}$ 
has non-zero $E2$ amplitudes for both $K=0$ and $K=2$ which are shown
in columns 8 and 10 respectively.

\begin{figure}[htb]
\centerline{\includegraphics[width=0.8\textwidth]{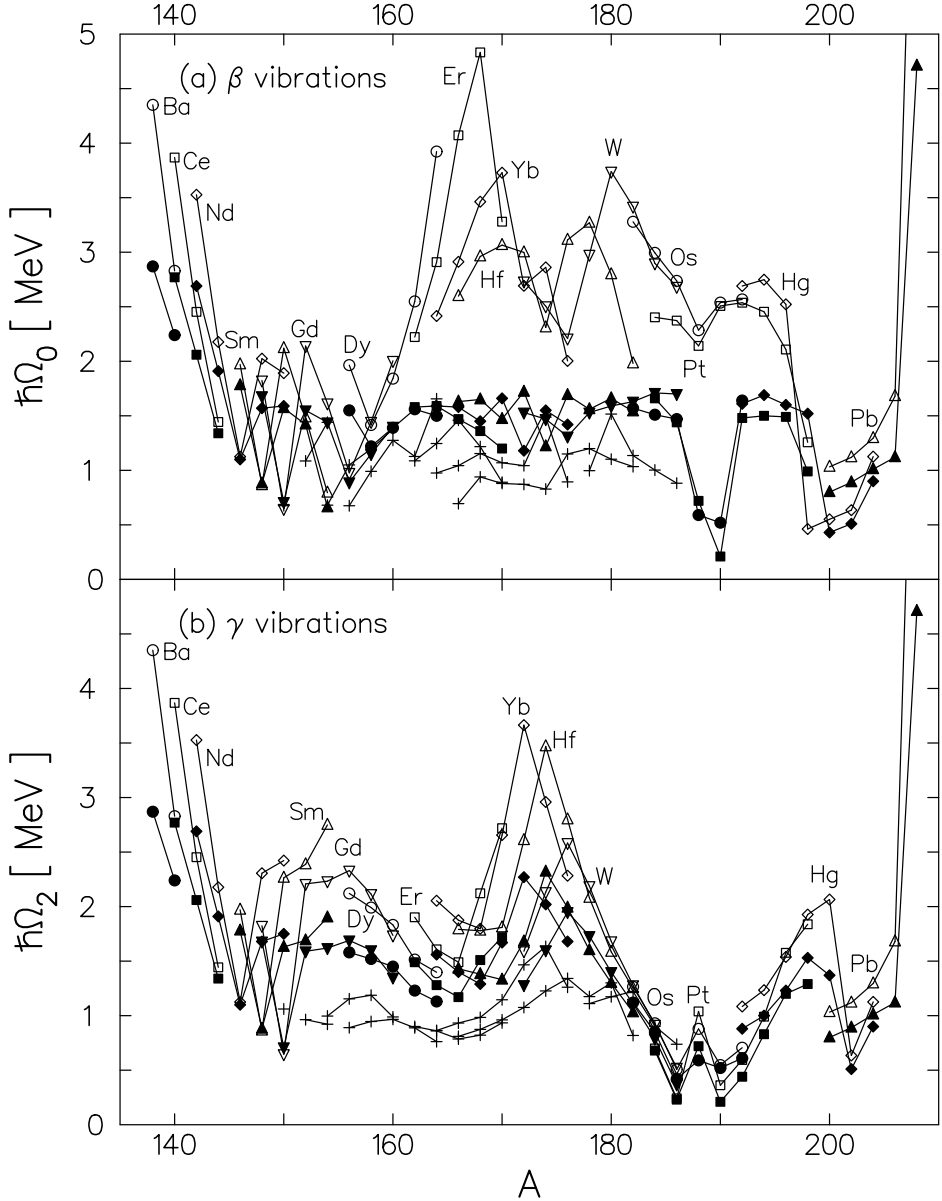}}
\caption{Calculated excitation energies in units of MeV, of $\beta$
(a) and $\gamma$ vibrations (b) as functions of mass number $A$ for
even-even nuclei in the rare-earth region.  The closed circles
indicate the RPA results while the open symbols are results of the
harmonic formula by Baranger and Kumar \protect\cite{BK_III}.  The
frequency of the harmonic formula for $^{208}$Pb is 10.9 MeV.  For
cases in which the HB ground states are spherical or triaxial, we
cannot distinguish $\beta$ from $\gamma$ vibrations and the same
values have been used for both (a) and (b).  Different symbols
indicate different isotopes; circles for Ba, Dy and Os; squares for
Ce, Er and Pt; diamonds for Nd, Yb, Hg; triangles (upwards) for Sm, Hf
and Pb; triangles (downward) for Gd and W. The experimental values are
represented by crosses for deformed nuclei \protect\cite{SHS91}.  }
\label{RPA_BK_omega}
\end{figure}

First we compare the RPA frequencies with the
harmonic formula given by Baranger and Kumar,
Eqs.~(13), (14) and (15) in Ref. \cite{BK_III}.
Unlike our results,
this formula contains a core contribution to the collective mass
(which is denoted by $B^{\rm c}$).
Since our RPA calculations are carried out in the same model space,
we should compare the RPA frequencies with those calculated
by setting $B^{\rm c}=0$.
These are shown in Fig.~\ref{RPA_BK_omega}.
Open symbols are the results of the harmonic formula while solid ones are
the RPA frequencies.
For spherical and triaxial nuclei, the same RPA frequencies are displayed
in the figure for $\beta$ and $\gamma$ vibrations.
The figure shows that the frequencies obtained by the formula are
larger than the RPA frequencies.
For $\gamma$ vibrations, the isotope dependence is well reproduced,
even though the absolute values are mostly overestimated by up to a factor two.
For $\beta$ vibrations,
the harmonic formula breaks down,
especially in well-deformed nuclei.
It fails to reproduce both the isotope dependence and absolute values.

Let us try to understand the origin of the harmonic formula.
It can be derived by taking the small-amplitude limit of
the adiabatic time-dependent theory of collective motion
developed by Baranger and Kumar \cite{BK_IV,BK_V}
which assumes that the collective coordinates are the expectation values
of the mass quadrupole operators
(more precisely the deformation parameters of the Nilsson potential).
In contrast with the formula,
the RPA normal-mode coordinates are self-consistently determined by
diagonalising the RPA matrix.
Therefore, the difference between open and closed symbols seen
in the figure can be attributed primarily to the difference
in the collective coordinates.

Although it is not the purpose of this paper to reproduce the
experimental data, we also show observed excitation energies of
$\beta$ and $\gamma$ bands for axially deformed nuclei ($Z=60\sim 74$)
in Fig.~\ref{RPA_BK_omega} \cite{SHS91}. Agreement with the experimental
data are significantly improved by choosing the collective coordinates
properly. One can see that the isotope dependence is roughly reproduced
in this model. However, we still overestimate the energies by typically
about 500 keV, which is at least partly due to the neglect of
anharmonic effects, and may point at the limitations of the simple model.

\begin{figure}[htb]
\centerline{\includegraphics[width=0.8\textwidth]{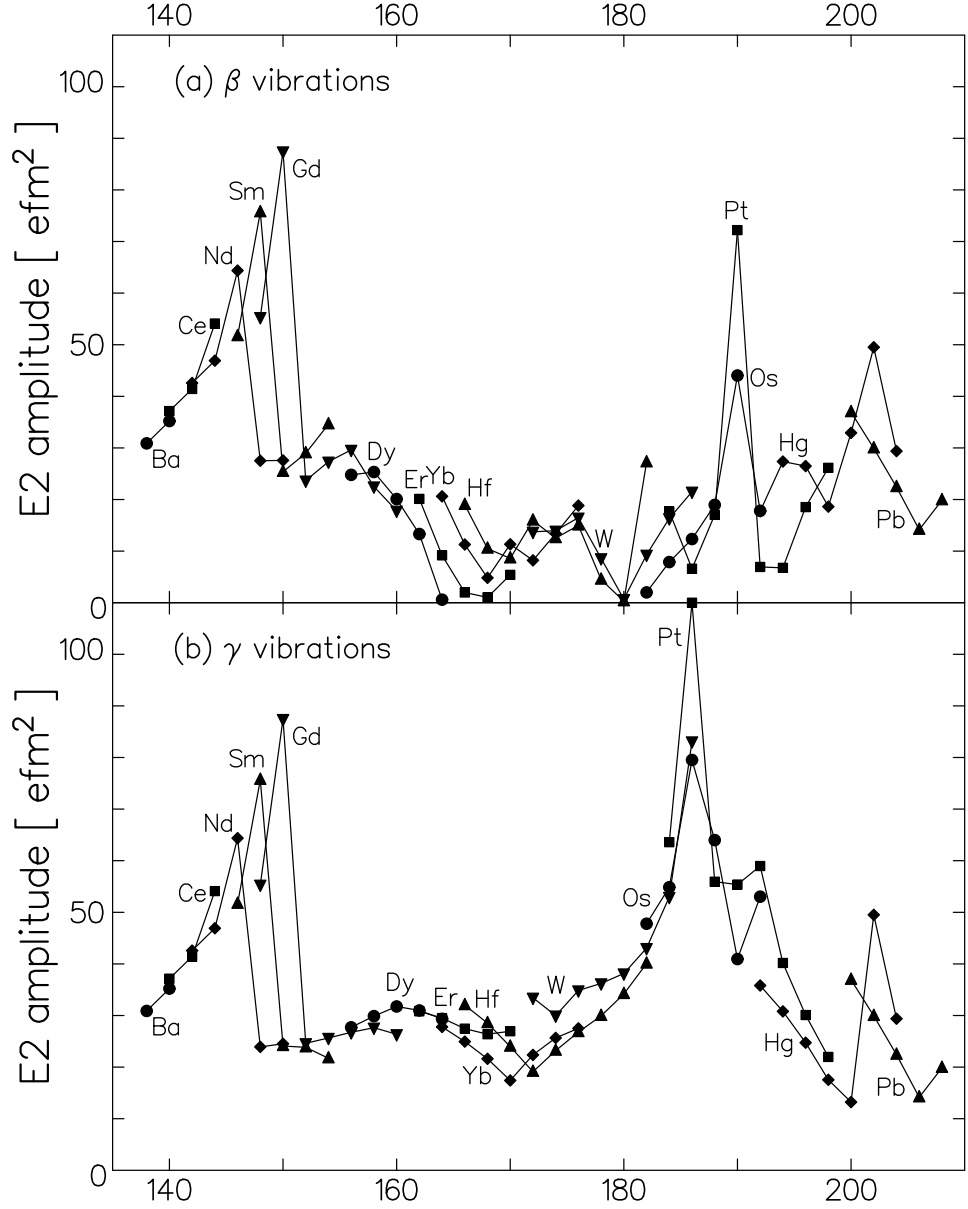}}
\caption{Calculated intrinsic $E2$ transition amplitudes
between the excited RPA and the ground state,
for $\beta$ (a) and $\gamma$ vibrations (b) in units of $e\mbox{fm}^2$.
For spherical nuclei, these values are also the $E2$ amplitudes
in laboratory frame.
For triaxial cases, a single state has both $K=0$ and $K=2$ components
of $E2$ amplitudes which are shown in (a) and (b), respectively.
See caption of Fig.~\ref{RPA_BK_omega} for symbols and the text for
further details.
}
\label{RPA_E2}
\end{figure}

In Fig.~\ref{RPA_E2}, we show $E2$ transition amplitudes for the RPA
states.
Evidently the $E2$ amplitudes are smaller for $\beta$ vibrations than
$\gamma$ vibrations.
This is partly because the $\beta$ vibrations are less collective, but
also because the collective coordinates for $\beta$ vibrations are
more complex than those for $\gamma$ motion (see below).

Now let us analyse the structure of the RPA normal-modes by
using the projection techniques described in Sec.~\ref{sec: PLHA},
using various sets of one-body operators.
First we utilise elementary multipole operators $\{ F^{(k)} \}$
on which the RPA matrix is projected.
We adopt monopole operators (rank-0),
quadrupole operators with radial dependence $r^0$ and $r^2$
(rank-2, $K=0,2$),
those with spin dependence (rank-2, $K=0,2$), and
hexadecapole operators (rank-4, $K=0,2$):
\begin{eqnarray}
\label{PRPA_op_1}
F^{(k)}=&& (P_+)_\tau, (P_-)_\tau,
   (Q_{20})_\tau, (Q_{22})_\tau,
   (r^0Y_{20})_\tau, (r^0Y_{22})_\tau, \nonumber\\
&& ([r^0Y_2 \times {\bf s} ]^{(2)}_{K=0})_\tau,
   ([r^0Y_2 \times {\bf s} ]^{(2)}_{K=2})_\tau,
   ([r^2Y_2 \times {\bf s} ]^{(2)}_{K=0})_\tau,
   ([r^2Y_2 \times {\bf s} ]^{(2)}_{K=2})_\tau, \nonumber \\
&& (r^4 Y_{40})_\tau, (r^4 Y_{42})_\tau , \quad \tau = n,p,
\end{eqnarray}
where all operators have the positive signature $r=+1$,
as in Eq.~(\ref{sig_good_Q}), but the superscript
$(+)$ denoting the  signature quantum number is suppressed from here on.
All matrix elements of rank-2 operators are multiplied by
factors $\zeta$ and $\alpha_\tau^2$,
just as for the quadrupole operators (\ref{modify_Q}).
For $\beta$ vibrations, there are fourteen relevant operators of $K=0$
while there are ten operators of $K=2$ for $\gamma$.
For a spherical case, the number reduces to eight (only rank-2 operators).
For a triaxial case, all 24 operators can mix together.

\begin{figure}[htb]
\centerline{\includegraphics[width=0.8\textwidth]{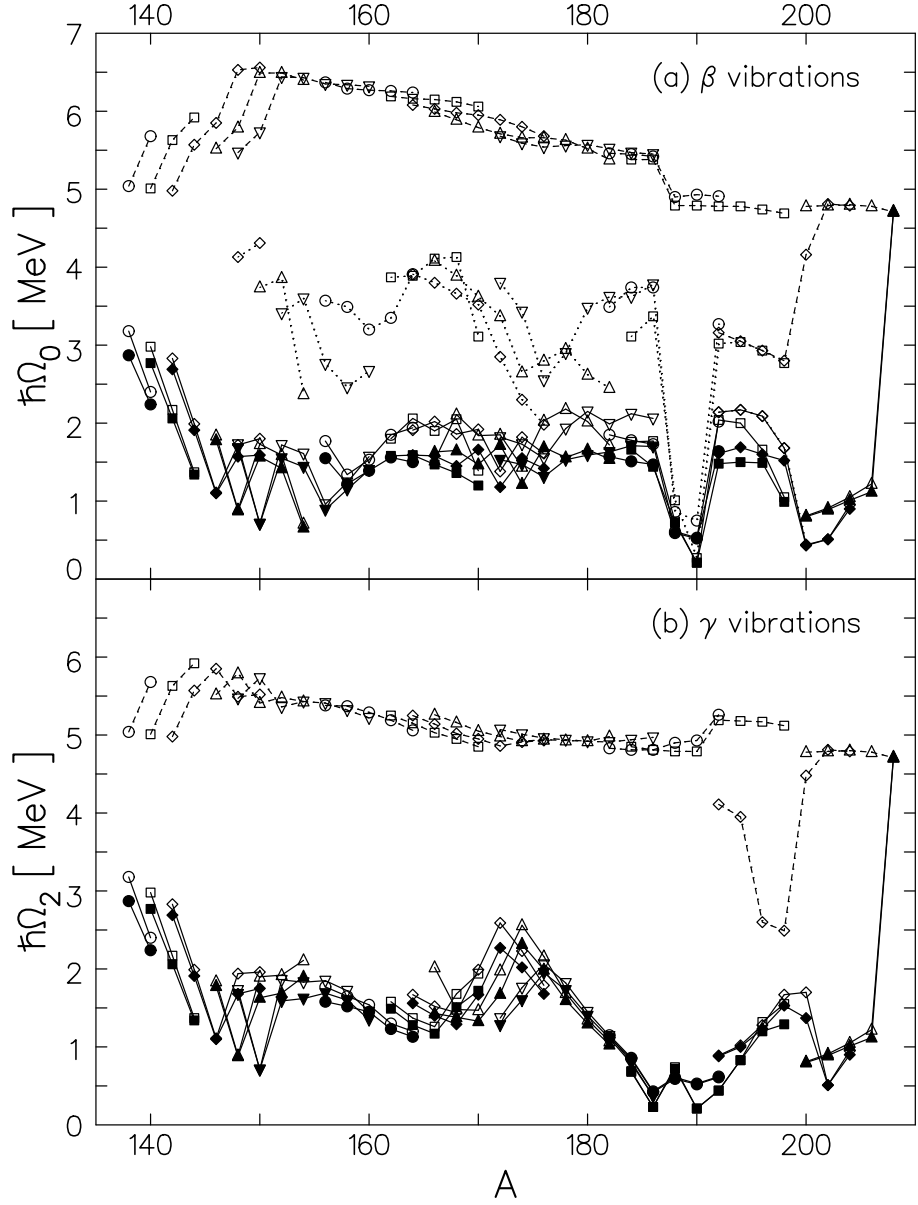}}
\caption{Calculated energies in units of MeV,
by means of both the projected RPA (open symbols)
and the exact RPA (closed) for $\beta$ (a) and $\gamma$ vibrations (b).
The results of projected RPA connected with dashed, dotted and solid lines
indicate different sets of one-body operators being employed
for the projection.
See the main text for further details.
}
\label{PRPA_omega}
\end{figure}

The calculated frequencies $\hbar\bar{\Omega}$ of the projected RPA equation
(\ref{projected_LHA}) 
are shown in Fig.~\ref{PRPA_omega} as
open symbols with dashed lines and are
compared with the real RPA frequencies (solid symbols).
As is clearly seen, the projection onto the operators (\ref{PRPA_op_1})
fails to reproduce
the excitation energies both for $\beta$ and $\gamma$ vibrations.
Not only the absolute values but also the isotope dependence turns out to be
incorrect for all nuclei (except for $^{208}$Pb).
The result indicates
that it is very difficult to obtain sensible approximation
by using elementary one-body operators.
This is mainly due to the fact that the projected RPA eigenvectors have
unrealistically large amplitudes for high-lying 2qp components.
In Ref. \cite{NWD99},
we have demonstrated for a few Sm isotopes that this can be remedied
by introducing a cut-off energy
for the 2qp matrix elements.
For $^{208}$Pb, which is spherical and has $\Delta_n=\Delta_p=0$,
the projection accidentally works well because
the set of available 1p1h modes is very limited,
$\nu(i_{11/2}(i_{13/2})^{-1})$, $\nu(g_{9/2}(i_{13/2})^{-1})$,
$\pi(h_{9/2}(h_{11/2})^{-1})$, $\pi(f_{7/2}(h_{11/2})^{-1})$,
all of which have almost the same particle-hole excitation energies.

In order to check the effect of suppressing the  high-energy components,
we also perform the projected RPA calculation
with the state-dependent operators whose 2qp matrix elements are
weighted with a factor $(E_{\rm 2qp})^{-2}$.
This means that we employ a set of hermitian
one-body operators $\{ \widetilde{F}^{(k)}\}$ defined by
\begin{equation}
\widetilde{F} \equiv \sum_\alpha
    \frac{F(\alpha)}{(E_\alpha)^2} (a^\dagger a^\dagger)_\alpha
    + \mbox{h.c.},
\label{scale}
\end{equation}
where $\alpha$ indicates the 2qp index and
$F(\alpha)=\bra{\alpha} F \ket{0}$.
This suppression factor $(E_{\rm 2qp})^{-2}$ can be derived analytically
as follows   
if the residual Hamiltonian consists only of a single-mode separable force
$H=-\frac{1}{2} \kappa R R$ and the RPA frequency is much smaller
than 2qp energies.
In the single-mode case,
from Eqs.~(\ref{B_PQ}) and (\ref{V_PQ}),
the RPA equation becomes
\begin{equation}
\left( (E_\alpha)^2 - (\hbar\Omega)^2 \right) f_{,\alpha} 
 = 2\kappa R(\alpha) \sum_\beta E_\beta R(\beta) f_{,\beta} .
\end{equation}
Thus, we can analytically determine the RPA
eigenvectors $f_\alpha$ as
\begin{equation}
f_{,\alpha} \propto \frac{R(\alpha)}{(E_\alpha)^2 - (\hbar\Omega)^2} .
\end{equation}
In the limit of $\hbar\Omega \ll E_\alpha$, this leads to the quoted
suppression factor $(E_{\rm 2qp})^{-2}$.

In the calculation, we adopt only the operators in the separable
interactions:
\begin{equation}
\label{PRPA_op_2}
\widetilde{F}^{(k)}=
   (\widetilde{P}_+)_\tau, (\widetilde{P}_-)_\tau,
   (\widetilde{Q}_{20})_\tau, (\widetilde{Q}_{22})_\tau,
   \quad \tau = n,p.
\end{equation}
This results in a six dimensional projected RPA matrix for $\beta$
vibrations and in two dimensional one for $\gamma$ vibrations and
for spherical cases.
Even for triaxial cases, the matrix dimension is thus only eight.
The results are shown in Fig.~\ref{PRPA_omega} by open symbols with
solid lines.
The results are dramatically improved from the previous ones and
the obtained frequencies are almost comparable to those of RPA
for both $\beta$ and $\gamma$ modes of all nuclei.
We can also see the significant improvement over results of
the harmonic formula, especially for the $\beta$
vibrations (Fig.~\ref{RPA_BK_omega}).
This is because we incorporate the monopole pairing operators in
Eq.~(\ref{PRPA_op_2}) in addition to the quadrupole ones.
It is worth noting that the frequency of the projected RPA gives an
upper limit to the real RPA frequency because it is obtained by solving
the RPA in a restricted space.
On the other hand, this is not necessarily true for the harmonic formula
of  Ref.~\cite{BK_III}.

Let us examine the eigenvectors of projected RPA equation in more detail.
For spherical nuclei, the RPA eigenvector has a good isoscalar character
and can be well approximated by
\begin{equation}
\bar{f}^Q = c_n (\widetilde{Q}_2)_n + c_p (\widetilde{Q}_2)_p ,
\end{equation}
with $c_n \approx c_p$.
Here the tilde indicates that the matrix elements
include a suppression factor as in Eq.~(\ref{scale}).
A violation of this isoscalar character is seen for
$^{138}$Ba, $^{140}$Ce, $^{142}$Nd and Pb nuclei
in which either the neutrons or protons pairing gap vanishes.
The component with finite pairing gap is found to be larger
than the one with zero gap,
so that $c_n > c_p$ for Pb and $c_n < c_p$ for
$^{138}$Ba, $^{140}$Ce and $^{142}$Nd.
For deformed nuclei, where the collectivity of the
vibrational states is smaller than for spherical nuclei and the pairing
modes can mix with the quadrupole ones, the situation is more complex.
In Tables~\ref{PRPA_Dy} and \ref{PRPA_Os},
taking Dy and Os isotopes as examples,
we show the coefficients of the approximate eigenvectors,
$\bar{f} = \sum_k c_k \widetilde{F}^{(k)}$,
with respect to the set of operators in Eq.~(\ref{PRPA_op_2}).
The $\gamma$ vibrations approximately have isoscalar character
and the collective coordinate can be approximated by
\begin{equation}
\bar{f}^\gamma = c_n (\widetilde{Q}_{22})_n + c_p (\widetilde{Q}_{22})_p ,
\end{equation}
with $c_n \approx c_p$.
For the $\beta$ vibration, no such simple form can be found, since
the properties of RPA eigenmodes change from one nucleus to the other.
The $\beta$ vibrations for $^{156-162}$Dy possess a quadrupole nature,
$c_n (\widetilde{Q}_{20})_n + c_p (\widetilde{Q}_{20})_p$,
with $c_n > c_p$,
although there are substantial contributions from the monopole pairing modes.
For $^{164}$Dy, there is a dominant pairing component $(P_+)_n$.
For Os isotopes, $^{182\sim 186}$Os have dominant proton pairing $(P_+)_p$,
while $^{192}$Os has a quadrupole nature ($c_n < c_p$).
For the triaxial nuclei $^{188,190}$Os,
although the $K=0$ and $K=2$ components are mixed,
the vibrational state is of
isoscalar quadrupole nature with $c_n \approx c_p$.
The changes observed in the RPA normal-mode coordinates
for the $\beta$
vibrations are difficult to be systematically summarised and
turn out to be very sensitive to details of the underlying shell structure.

In order to confirm the importance of the monopole components
for $\beta$ vibrations,
we have performed a calculation using only the quadrupole operators
weighted with the factor $E_{\rm 2qp}^{-2}$,
$(\widetilde{Q}_{20})_\tau$ and $(\widetilde{Q}_{22})_\tau$
($\tau=n,p$).
For spherical nuclei, there is no change from the previous results
since the monopole and quadrupole components are exactly decoupled.
For deformed nuclei only, the results are shown in Fig.~\ref{PRPA_omega}
by open symbols with dotted lines.
Differences between solid and dotted lines come from the contributions
of monopole pairing operators.
The figure clearly shows that the normal-mode coordinate for $\beta$
vibrations has a significant amount of monopole components.

We now wish to quantify the quality of the approximation.
One measure is the difference between 
the real RPA frequencies $\hbar\Omega$ and
the projected RPA $\hbar\bar{\Omega}$,
but as usual measures based on the eigenvectors are more sensitive.
We define two criteria to directly check the closeness between
the eigenvectors $f$ and $\bar{f}$.
Choosing the normalisation (of $f$ and $\bar{f}$)
$f_{,\alpha} B^{\alpha\beta} f_{,\beta} =
\bar{f}_{,\alpha} B^{\alpha\beta} \bar{f}_{,\beta} = 1$,
those are calculated as
\begin{eqnarray}
\label{delta_B}
\delta_B &=&
   f_{,\alpha} B^{\alpha\beta} ( f_{,\beta}-\bar{f}_{,\beta} )
 =1 - f_{,\alpha} B^{\alpha\beta} \bar{f}_{,\beta} ,\\
\delta_1 &=& 1-\frac{(f,\bar{f})}
                         {\sqrt{(f, f)(\bar{f},\bar{f})}} ,
\end{eqnarray}
where $(f,f')=\sum_\alpha f_{,\alpha} f'_{,\alpha}$.
We have $0 \leq \delta \leq 1$ with
$\delta=0$ corresponding to the exact projection and $\delta=1$ to the
case where $\bar{f}$ is orthogonal to $f$.
The values of $\delta_B$ and $\delta_1$ are listed in
Tables~\ref{PRPA_Dy} and \ref{PRPA_Os}.
From the values of $\delta_B$,
we see that the projection works reasonably well
for Dy and Os isotopes.
The worst case is the $\beta$ vibration of $^{192}$Os for which
$\delta_B\approx 0.3$.
Roughly speaking, the one-body operators (\ref{PRPA_op_2}) possess
70\% of overlap with the real eigenvectors for $^{192}$Os
($\delta_B\approx 0.3$),
and more than 90\% for the others ($\delta_B \lesssim 0.1$).

The values of $\delta_1$ for $\beta$ vibrations are often much larger than
the corresponding $\delta_B$
(see $^{156,158}$Dy and triaxial nuclei $^{188,190}$Os).
This indicates that the $\bar{f}$ for $\beta$ contains some spurious
components.
In the present calculations,
the pairing and spatial rotations are the obvious spurious motions
which are zero eigenmodes of the mass matrix,
\begin{equation}
\label{PQ_spurious}
B^{\alpha\beta} f^I_{,\beta} = 0 , \quad\mbox{for}\quad
f^I = N_n, N_p, J_x .
\end{equation}
The measure $\delta_B$ (\ref{delta_B}) is insensitive to admixture of these
spurious components in $\bar{f}$.
On the other hand, the value of $\delta_1$ will increase by mixing in
spurious components.
As we have discussed in the last part of Sec.~\ref{sec: PLHA},
the spurious modes $f^I$ do not affect the projected LHA equations.
Therefore, the large values of $\delta_1$ should not cause serious problems
in application of the projected LHA to the large amplitude collective motion.

\begin{figure}[htb]
\centerline{\includegraphics[width=0.8\textwidth]{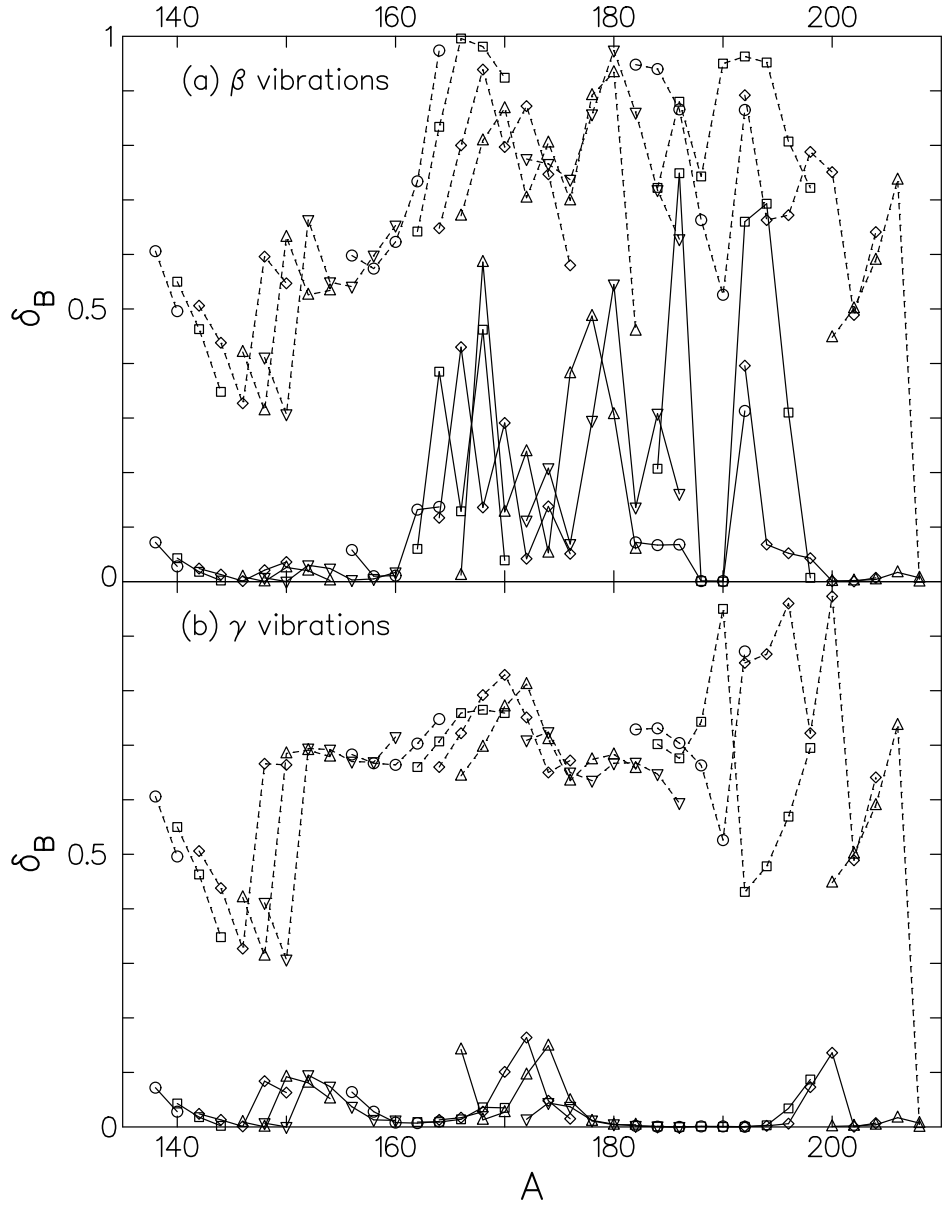}}
\caption{Quality of projection $\delta_B$, Eq.~(\ref{delta_B}),
for $\beta$ (a) and $\gamma$ vibrations (b).
Dashed lines and solid lines denote
different sets of one-body operators being employed for the projection.
See the main text for further details.
}
\label{PRPA_delta}
\end{figure}

In Fig.~\ref{PRPA_delta}, the values of $\delta_B$ are shown for
the projected RPA using the set of elementary operators (\ref{PRPA_op_1})
(represented by dashed lines) and the set of state-dependent operators
containing the factor $E_{\rm 2qp}^{-2}$ (\ref{PRPA_op_2}) (solid lines).
For $\gamma$ vibrations, the suppression factor $E_{\rm 2qp}^{-2}$
can improve the quality of projection by an order of magnitude
and we have $\delta_B \lesssim 0.1$ for all nuclei.
For $\beta$ vibrations, there are some nuclei in the range $Z=68\sim 80$
for which the operators in Eq.~(\ref{PRPA_op_2}) are not enough to produce
the good approximation.
This confirms the complexity of the normal-mode coordinates for $\beta$.

\begin{figure}[htb]
\centerline{\includegraphics[width=0.4\textwidth]{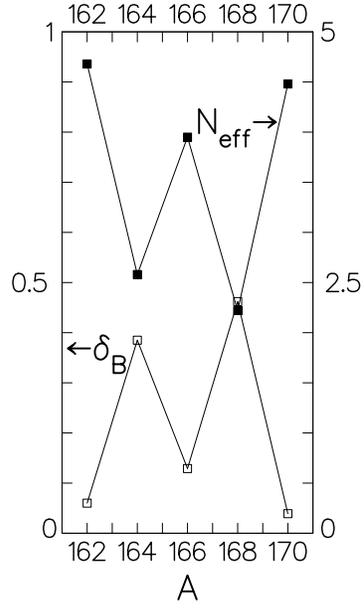}}
\caption{The measure $\delta_B$ (open squares and scale on the left)
and the measure of collectivity $N_{\rm eff}$, Eq.~(\ref{Neff}),
(closed squares and scale on the right) for $\beta$ vibrations in the
Er isotopes.
}
\label{PRPA_stag}
\end{figure}

We also see a strong isotope dependence in Fig.~\ref{PRPA_delta}a.
For instance, the Er isotopes show a staggering behaviour in $\delta_B$.
This behaviour is found to be
closely related to the collectivity of the states.
Since the $\beta$ vibrations contains the pairing collectivity in addition to
the quadrupole one,
the $E2$ amplitudes are not necessarily the best indicator of the
collectivity of the states.
We show in Fig.~\ref{PRPA_stag} a following quantity for the RPA eigenmodes
of $\beta$ vibrations in Er nuclei \cite{AKL81}:
\begin{equation}
\label{Neff}
N_{\rm eff}^\mu \equiv 
 \frac{\sum_\alpha ( \hbar\Omega f^\mu_{,\alpha}f^\mu_{,\alpha}
                     + (\hbar\Omega)^{-1} g^\alpha_{,\mu} g^\alpha_{,\mu} )}
  {\sum_\beta f^\mu_{,\beta} g^\beta_{,\mu}
                        (\hbar\Omega f^\mu_{,\beta}f^\mu_{,\beta}
                        +(\hbar\Omega)^{-1} g^\beta_{,\mu} g^\beta_{,\mu})} ,
\quad \mbox{(no summation with respect to $\mu$).}
\end{equation}
If $n$ 2qp components equally contribute to the RPA mode, we have
$N_{\rm eff}=n$.
Thus, Eq. (\ref{Neff}) gives an effective number of 2qp excitations
of which the RPA eigenmode consists.
The figure shows that the collectivity of $\beta$ vibrations in Er isotopes
is weak (compared to the $\gamma$ vibrations for which
$N_{\rm eff} \gtrsim 8$ in most cases),
and $N_{\rm eff}$ shows a staggering pattern correlated with that for
$\delta_B$ (while the $E2$ amplitudes in Fig.~\ref{RPA_E2} show no sign of
staggering behaviour).
For $^{164,168}$Er, there are dominant 2qp components in the RPA eigenmodes,
which decrease $N_{\rm eff}$ and increase $\delta_B$.
For those cases where $\delta_B \gtrsim 0.3$ even with state-dependent
operators, the $\beta$ vibrations are not collective at all,
$N_{\rm eff} \lesssim 3$.

\section{conclusion and discussion}
\label{sec: conclusions}

We have examined different version of LHA techniques
to see whether the conservation laws in the quantum level are
satisfied in the classical LHA.
It turns out that the restriction to point transformations must
be lifted in order that spurious motion leads to the exact zero modes.
The symplectic version of the LHA, which includes an extended adiabatic
approximation,
can guarantee the exact separation of zero-energy spurious modes.

The projected LHA can be used to truncate the local RPA calculations.
We have shown that there is no problem with spurious modes 
for the projection method, at least
when the spurious modes are zero eigenmodes of the mass matrix.
This is exactly the case for the pairing and spatial rotation in the
P+Q model for which
we have investigated the structure of the collective coordinates
for quadrupole vibrations
and examined the possibility of expressing the self-consistent
cranking operator in terms of a limited number of one-body operators.
It seems very difficult to approximate the normal-mode vectors
with  only elementary one-body operators.
The difficulty disappears, however,
when we use a small number of {\it state-dependent} one-body operators.
At HB equilibrium, the collective coordinates for quadrupole vibrations
in spherical nuclei and for $\gamma$ vibrations ($K=2$) in axially deformed
nuclei can be roughly approximated as $f\approx \widetilde{Q}_{2K}$, where
$\widetilde{Q}_{2K}$ is a state-dependent ``mass'' quadrupole operator
defined by Eq.~(\ref{modify_Q}) weighted by a factor
$E_{\rm 2qp}^{-2}$ for each 2qp matrix element (\ref{scale}).
However, for $\beta$ vibrations in axial deformed nuclei and 
for all excitations in triaxial nuclei,
the monopole pairing components are as important as the quadrupole ones.
It is worth emphasising that the collective coordinates are
very different from the usual (state-independent) mass quadrupole
operator even for $\gamma$ and spherical cases.
This shows the importance of a self-consistent determination of
the collective coordinates for large amplitude collective motion,
because the coordinates are now found to have a strong state-dependence as well.
The structure of the self-consistent cranking operators is clearly
changing when we move from spherical to axially deformed and triaxial
nuclei.
For the study of large amplitude
collective motion in heavy nuclei for which
the diagonalisation of the RPA matrix becomes too time-consuming,
the results of this paper gives a suggestion for a
choice of a state-dependent basis of operators.
The choice of a limited set of (state-dependent) basis operators
provides a practical way to solve the LHA through the projection.
With the use of self-consistent cranking operators,
the LHA should provide a significant improvement over a conventional
CHFB calculation based on  fixed cranking operators.

Despite all the above qualities,    
the symplectic version of LHA has a difficulty, namely  
the determination of $f^\mu_{,\alpha\beta}$,
the second derivatives of the final
coordinates $q^\mu$ with respect to the original coordinates $\xi^\alpha$.
For the extended adiabatic transformation (\ref{EAT_0}),
we also need to calculate $f^{(1)\mu\alpha\beta}$,
the terms of second order in $\pi$.
At HB states, we need these quantities for spurious coordinates $\mu=I$ only,
$f^I_{,\alpha\beta}$ and $f^{(1)I\alpha\beta}$,
which can be calculated
because we know {\it a priori} the one-body operators corresponding to
the spurious motions.
We have done this for P+Q model.
Away from equilibrium, however, in addition to these $\mu=I$,
we need the second derivatives of the collective coordinate, $\mu=1$.
This is not trivial because
the local RPA equation determines only the first derivatives,
$f^1_{,\alpha}$ and $g^\alpha_{,1}$.
Unfortunately, so far we do not have a general method to
determine $f^1_{,\alpha\beta}$ and $f^{(1)1\alpha\beta}$.
The projection method may actually provide us with a partial answer
\cite{WKD90}.
If we project the LHA equation onto a set of one-body operators
$\{ F^{(k)} \}$,
the equation provides us with a local representation of
collective coordinate in terms of a linear combination of the $F^{(k)}$'s.
Using the explicit form
\begin{equation}
F^{(k)} = \sum_{i > j} F^{(k)}(ij)
                          (a_i^\dagger a_j^\dagger + a_j a_i )
         + \sum_{ij} F^{'(k)}(ij) a_i^\dagger a_j ,
\end{equation}
we find
\begin{equation}
\label{f_alpha}
f^1_{,\alpha} \equiv \frac{\partial q^1}{\partial \xi^\alpha}
 = \sum_k c_k {\cal F}^{(k)}_{,\alpha}
 = \sqrt{2} \sum_k c_k F^{(k)}(\alpha) ,
\end{equation}
where $F^{(k)}(\alpha)=F^{(k)}(ij)$ for 2qp index $\alpha=(ij)$ and
we assume that the operators $F^{(k)}$ are hermitian.
Here we use a local coordinate system, in which the point
under consideration is the origin of the phase space,
and the relation (\ref{2qp_R}) to obtain Eq.~(\ref{f_alpha}).
Since we know the explicit form of the operators $F^{(k)}$,
we are now able to calculate the second derivatives $f^1_{,\alpha\beta}$ and
second-order terms in momentum $f^{(1)1\alpha\beta}$.
\begin{eqnarray}
\label{f_ab}
f^1_{,\alpha\beta} &\equiv&
 \frac{\partial^2 q^1}{\partial\xi^\alpha\partial\xi^\beta}
 \approx \sum_k c_k {\cal F}^{(k)}_{,\alpha\beta}
 = 4 \sum_k c_k G^{(k)}(\alpha\beta) ,\\
\label{f_1ab}
f^{(1)1\alpha\beta} &\equiv&
 \frac{\partial^2 q^1}{\partial\pi_\alpha\partial\pi_\beta}
 \approx \sum_k c_k \frac{\partial^2{\cal F}^{(k)}}
                     {\partial\pi_\alpha\partial\pi_\beta}
 = 4 \sum_k c_k G^{(k)}(\alpha\beta) ,
\end{eqnarray}
where we neglect the derivative of $c_k$, for simplicity.
$G^{(k)}(\alpha\beta)$ are related to $F^{'(k)}(ij)$, by
\begin{equation}
G^{(k)}(\alpha\beta) \equiv F^{'(k)}(ij) \delta_{kl} , \quad
 \mbox{for }\alpha=(ik), \beta=(jl).
\end{equation}
Although, following this procedure, it is possible to calculate
$f^1_{,\alpha\beta}$ and $f^{(1)1\alpha\beta}$,
we have to note that, unlike the determination of $f^1_{,\alpha}$ by means
of the RPA,
the determination of these quantities through Eqs.~(\ref{f_ab}) and
(\ref{f_1ab}) does not depend on the dynamics of the system.
In this approach, part of the dynamics is hidden in the coordinate dependence of the
coefficients $c_k$, which we have ignored for this simple discussion.

In this paper,
we have mainly discussed the structure of the self-consistent 
cranking operators for the P+Q model only at HB minimum points.
The structure at points away from the minimum
and the further application of the LHA technique
will be the subject of a future publication.

\acknowledgments

This work was supported by a research grant (GR/L22331) from
the Engineering and Physical Sciences Research Council (EPSRC)
of Great Britain, and through a grant (PN 98.044) from
ALLIANCE, the Franco-British Joint Research programme.
The Laboratoire de Physique Th\'eorique is a Unit\'e Mixte de Recherche    
du C.N.R.S., UMR 8627.
We thank Y.~R.~Shimizu for providing us with a numerical code of the RPA
calculation for the pairing-plus-quadrupole forces.


\newcommand\s{\ \ \ \ }
\begin{table}[htb]
\caption{
HB equilibrium shapes ($\beta,\gamma$), pairing gaps $\Delta_\tau$,
$\tau=n,p$ (in units of MeV),
RPA frequencies $\hbar\Omega_K$ (in MeV) and
$E2$ transition amplitudes $M(E2)_K$ (in $e\mbox{fm}^2$).
The RPA results for spherical nuclei are shown in columns 7 and 8
(columns 9 and 10 are left blank),
while for triaxial ones the RPA results are in columns 7, 9 and 10
(column 8 is blank).
Entries labelled with asterisks indicate that they do not correspond to   
the lowest excited states.
See the main text for further details.
}
\begin{tabular}{crrrrrrrrrr}
& $Z$ & $N$ & $\beta$ & $\gamma$ & $\Delta_n$ & $\Delta_p$ & $\hbar\Omega_0$
 & $M(E2)_0$ & $\hbar\Omega_2$ & $M(E2)_2$ \\
\hline
Ba &  56  & 82 & 0.0\s & 0.0 & 0.0\s & 1.557 & 2.87 & 30.88\\ 
   &      & 84 & 0.0\s & 0.0 & 0.913 & 1.528 & 2.24 & 35.21\\ 
Ce &  58  & 82 & 0.0\s & 0.0 & 0.0\s & 1.686 & 2.77 & 37.14\\ 
   &      & 84 & 0.0\s & 0.0 & 0.895 & 1.654 & 2.06 & 41.43\\ 
   &      & 86 & 0.0\s & 0.0 & 1.198 & 1.624 & 1.34 & 54.11\\ 
Nd &  60  & 82 & 0.0\s & 0.0 & 0.0\s & 1.766 & 2.69 & 42.58\\ 
   &      & 84 & 0.0\s & 0.0 & 0.878 & 1.734 & 1.91 & 46.91\\ 
   &      & 86 & 0.0\s & 0.0 & 1.176 & 1.703 & 1.10 & 64.37\\ 
   &      & 88 & 0.184 & 0.0 & 0.964 & 1.386 & 1.57 & 27.51 & 1.68 & 23.92\\ 
   &      & 90 & 0.236 & 0.0 & 0.981 & 1.229 & 1.59 & 27.58 & 1.75 & 24.52\\ 
Sm &  62  & 84 & 0.0\s & 0.0 & 0.861 & 1.778 & 1.78 & 51.66\\ 
   &      & 86 & 0.0\s & 0.0 & 1.155 & 1.746 & 0.88 & 75.68\\ 
   &      & 88 & 0.198 & 0.0 & 0.910 & 1.394 & 1.57 & 25.37 & 1.63 & 24.09\\ 
   &      & 90 & 0.247 & 0.0 & 0.944 & 1.228 & 1.42 & 28.97 & 1.69 & 23.83\\ 
   &      & 92 & 0.312 & 0.0 & 0.902 & 1.007 & 0.66 & 34.58 & 1.90 & 21.71\\ 
Gd &  64  & 84 & 0.0\s & 0.0 & 0.845 & 1.793 & 1.68 & 55.34\\ 
   &      & 86 & 0.0\s & 0.0 & 1.134 & 1.761 & 0.71 & 87.45\\ 
   &      & 88 & 0.203 & 0.0 & 0.873 & 1.386 & 1.55 & 23.66 & 1.59 & 24.61\\ 
   &      & 90 & 0.249 & 0.0 & 0.917 & 1.221 & 1.44 & 27.33 & 1.62 & 25.65\\ 
   &      & 92 & 0.299 & 0.0 & 0.904 & 1.056 & 0.89 & 29.62 & 1.69 & 26.74\\ 
   &      & 94 & 0.320 & 0.0 & 0.880 & 0.980 & 1.15 & 22.52 & 1.60 & 27.64\\ 
   &      & 96 & 0.329 & 0.0 & 0.840 & 0.938 & 1.40 & 17.82 & 1.35 & 26.36\\ 
Dy &  66  & 90 & 0.244 & 0.0 & 0.897 & 1.212 & 1.55 & 24.78 & 1.58 & 27.71\\ 
   &      & 92 & 0.285 & 0.0 & 0.901 & 1.069 & 1.22 & 25.33 & 1.52 & 29.88\\ 
   &      & 94 & 0.308 & 0.0 & 0.873 & 0.988 & 1.39 & 20.09 & 1.45 & 31.75\\ 
   &      & 96 & 0.320 & 0.0 & 0.827 & 0.941 & 1.56 & 13.31 & 1.23 & 30.98\\ 
   &      & 98 & 0.327 & 0.0 & 0.748 & 0.904 & 1.50 &  0.59 & 1.13 & 29.35\\ 
Er &  68  & 94 & 0.295 & 0.0 & 0.867 & 0.986 & 1.58 & 20.12 & 1.49 & 30.81\\ 
   &      & 96 & 0.310 & 0.0 & 0.817 & 0.927 & 1.59 &  9.18 & 1.28 & 29.50\\ 
   &      & 98 & 0.320 & 0.0 & 0.736 & 0.884 & 1.47 &  1.98 & 1.17 & 27.45\\ 
   &      &100 & 0.326 & 0.0 & 0.680 & 0.849 & 1.36 &  1.03 & 1.51 & 26.37\\ 
   &      &102 & 0.330 & 0.0 & 0.624 & 0.819 & 1.20 &  5.40 & 1.72 & 26.95\\ 
Yb &  70  & 94 & 0.278 & 0.0 & 0.874 & 1.016 & 1.59 & 20.62 & 1.56 & 27.79\\ 
   &      & 96 & 0.296 & 0.0 & 0.817 & 0.946 & 1.58 & 11.28 & 1.40 & 24.94\\ 
   &      & 98 & 0.309 & 0.0 & 0.734 & 0.892 & 1.45 &  4.82 & 1.29 & 21.61\\ 
   &      &100 & 0.316 & 0.0 & 0.679 & 0.852 &*1.66 &*11.33 & 1.67 & 17.41\\ 
   &      &102 & 0.320 & 0.0 & 0.632 & 0.820 & 1.18 &  8.21 &*2.27 &*22.36\\ 
   &      &104 & 0.313 & 0.0 & 0.739 & 0.813 &*1.55 &*13.39 & 2.02 & 25.66\\ 
   &      &106 & 0.305 & 0.0 & 0.803 & 0.811 & 1.42 & 18.82 & 1.68 & 27.51\\ 
Hf &  72  & 94 & 0.252 & 0.0 & 0.907 & 1.060 & 1.63 & 18.96 & 1.43 & 32.01\\ 
   &      & 96 & 0.274 & 0.0 & 0.842 & 1.004 & 1.65 & 10.49 & 1.38 & 28.52\\ 
   &      & 98 & 0.289 & 0.0 & 0.759 & 0.963 & 1.47 &  8.59 & 1.33 & 23.97\\ 
   &      &100 & 0.299 & 0.0 & 0.706 & 0.932 &*1.72 &*15.93 & 1.68 & 19.10\\ 
   &      &102 & 0.302 & 0.0 & 0.678 & 0.907 & 1.22 & 12.52 &*2.32 &*23.19\\ 
   &      &104 & 0.293 & 0.0 & 0.744 & 0.893 &*1.69 &*14.97 & 1.99 & 26.79\\ 
   &      &106 & 0.282 & 0.0 & 0.785 & 0.884 & 1.56 &  4.48 & 1.60 & 29.93\\ 
   &      &108 & 0.272 & 0.0 & 0.824 & 0.878 & 1.66 &  0.28 & 1.30 & 34.17\\ 
   &      &110 & 0.257 & 0.0 & 0.920 & 0.883 & 1.54 & 27.21 & 1.03 & 40.07\\ 
W  &  74  & 98 & 0.264 & 0.0 & 0.819 & 0.962 & 1.53 & 13.74 & 1.28 & 33.44\\ 
   &      &100 & 0.275 & 0.0 & 0.768 & 0.940 & 1.47 & 13.95 & 1.60 & 29.98\\ 
   &      &102 & 0.277 & 0.0 & 0.752 & 0.918 & 1.31 & 16.48 & 1.95 & 34.85\\ 
   &      &104 & 0.270 & 0.0 & 0.777 & 0.893 & 1.53 &  8.60 & 1.73 & 36.22\\ 
   &      &106 & 0.259 & 0.0 & 0.796 & 0.872 & 1.59 &  0.68 & 1.40 & 38.14\\ 
   &      &108 & 0.247 & 0.0 & 0.838 & 0.856 & 1.63 &  9.29 & 1.11 & 43.06\\ 
   &      &110 & 0.228 & 0.0 & 0.932 & 0.854 & 1.71 & 16.35 & 0.79 & 52.97\\ 
   &      &112 & 0.202 & 0.0 & 0.988 & 0.872 & 1.70 & 21.51 & 0.37 & 83.11\\ 
Os &  76  &106 & 0.235 & 0.0 & 0.838 & 0.785 & 1.57 &  2.00 & 1.12 & 47.79\\ 
   &      &108 & 0.222 & 0.0 & 0.883 & 0.762 & 1.51 &  7.88 & 0.85 & 54.86\\ 
   &      &110 & 0.201 & 0.0 & 0.966 & 0.758 & 1.47 & 12.34 & 0.42 & 79.52\\ 
   &      &112 & 0.181 &22.1 & 0.990 & 0.789 & 0.59 & 18.96 &      & 63.99\\ 
   &      &114 & 0.171 &46.0 & 0.982 & 0.830 & 0.52 & 44.03 &      & 40.94\\ 
   &      &116 &$-$0.156 & 0.0 & 0.928 & 0.844 & 1.64 & 17.81 & 0.61 & 53.03\\ 
Pt &  78  &106 & 0.205 & 0.0 & 0.930 & 0.754 &*1.66 &*17.68 & 0.68 & 63.59\\ 
   &      &108 & 0.190 & 0.0 & 0.976 & 0.723 & 1.44 &  6.59 & 0.23 & 110.48\\
   &      &110 & 0.171 &20.1 & 1.046 & 0.671 & 0.72 & 17.05 &      & 55.93\\ 
   &      &112 & 0.160 &52.6 & 1.078 & 0.674 & 0.21 & 72.21 &      & 55.38\\ 
   &      &114 &$-$0.149 & 0.0 & 1.030 & 0.676 & 1.44 &  6.93 & 0.44 & 58.97\\ 
   &      &116 &$-$0.138 & 0.0 & 0.944 & 0.676 & 1.50 &  6.76 & 0.83 & 40.18\\ 
   &      &118 &$-$0.126 & 0.0 & 0.814 & 0.678 & 1.49 & 18.53 & 1.20 & 30.09\\ 
   &      &120 &$-$0.107 & 0.0 & 0.693 & 0.692 & 0.99 & 26.14 & 1.29 & 21.97\\ 
Hg &  80  &112 &$-$0.139 & 0.0 & 1.106 & 0.0\s & 1.61 & 17.73 & 0.88 & 35.81\\ 
   &      &114 &$-$0.131 & 0.0 & 1.045 & 0.0\s & 1.69 & 27.36 & 1.00 & 30.80\\ 
   &      &116 &$-$0.123 & 0.0 & 0.956 & 0.0\s & 1.60 & 26.48 & 1.23 & 24.73\\ 
   &      &118 &$-$0.114 & 0.0 & 0.830 & 0.0\s & 1.52 & 18.62 & 1.53 & 17.54\\ 
   &      &120 &$-$0.089 & 0.0 & 0.745 & 0.349 & 0.43 & 32.92 & 1.37 & 13.22\\ 
   &      &122 & 0.0\s & 0.0 & 0.749 & 0.606 & 0.51 & 49.51\\ 
   &      &124 & 0.0\s & 0.0 & 0.510 & 0.596 & 0.90 & 29.38\\ 
Pb &  82  &118 & 0.0\s & 0.0 & 1.011 & 0.0\s & 0.80 & 36.93\\ 
   &      &120 & 0.0\s & 0.0 & 0.894 & 0.0\s & 0.89 & 29.92\\ 
   &      &122 & 0.0\s & 0.0 & 0.734 & 0.0\s & 1.01 & 22.41\\ 
   &      &124 & 0.0\s & 0.0 & 0.497 & 0.0\s & 1.12 & 14.14\\ 
   &      &126 & 0.0\s & 0.0 & 0.000 & 0.0\s & 4.71 & 19.86\\ 
\end{tabular}
\label{HFB_plus_RPA}
\end{table}

\renewcommand\s{\ \ \ }
\begin{table}[htb]
\caption{
RPA frequencies $\hbar\Omega$, projected RPA frequencies $\hbar\bar{\Omega}$,
(both in units of MeV)
and projected RPA solutions
for $\beta$ and $\gamma$ vibrations in Dy isotopes.
Coefficients for the approximated RPA normal-mode coordinates, $c_k$ of
$\bar{f}_{,\alpha} = \sum_k c_k \widetilde{F}^{(k)}(\alpha)$,
are listed normalized as $\sum_k |c_k|^2 = 1$.
The two possible measures of the quality of projection, 
$\delta_B$ and $\delta_1$, are shown at the bottom.
}
\begin{tabular}{crrrrrrrrrr}
&\multicolumn{2}{c}{$^{156}$Dy} & \multicolumn{2}{c}{$^{158}$Dy} &
\multicolumn{2}{c}{$^{160}$Dy} & \multicolumn{2}{c}{$^{162}$Dy} &
\multicolumn{2}{c}{$^{164}$Dy} \\
&\multicolumn{1}{c}{$\beta$} & \multicolumn{1}{c}{$\gamma$} &
 \multicolumn{1}{c}{$\beta$} & \multicolumn{1}{c}{$\gamma$} &
 \multicolumn{1}{c}{$\beta$} & \multicolumn{1}{c}{$\gamma$} &
 \multicolumn{1}{c}{$\beta$} & \multicolumn{1}{c}{$\gamma$} &
 \multicolumn{1}{c}{$\beta$} & \multicolumn{1}{c}{$\gamma$} \\
\hline
$\hbar\Omega$& \multicolumn{1}{c}{1.55}& \multicolumn{1}{c}{1.58}&
 \multicolumn{1}{c}{1.22}& \multicolumn{1}{c}{1.52}& \multicolumn{1}{c}{1.39}&
 \multicolumn{1}{c}{1.45}& \multicolumn{1}{c}{1.56}& \multicolumn{1}{c}{1.23}&
 \multicolumn{1}{c}{1.50}& \multicolumn{1}{c}{1.13}\\
$\hbar\bar{\Omega}$&\multicolumn{1}{c}{1.77}& \multicolumn{1}{c}{1.77}&
 \multicolumn{1}{c}{1.34}& \multicolumn{1}{c}{1.65}& \multicolumn{1}{c}{1.53}&
 \multicolumn{1}{c}{1.54}& \multicolumn{1}{c}{1.85}& \multicolumn{1}{c}{1.30}&
 \multicolumn{1}{c}{1.94}& \multicolumn{1}{c}{1.20}\\
\hline
$(\widetilde{P}_+)_n$ &
$-0.061$ &$0.0\s$ &$-0.071$ &$0.0\s$ &$ 0.030$ &$0.0\s$ &$ 0.106$ &$0.0\s$ &
  $ 0.988$ &$0.0\s$\\
$(\widetilde{P}_-)_n$ &
$-0.494$ &$0.0\s$ &$-0.437$ &$0.0\s$ &$-0.350$ &$0.0\s$ &$-0.312$ &$0.0\s$ &
  $ 0.064$ &$0.0\s$\\
$(\widetilde{Q}_{20})_n$ &
$ 0.609$ &$0.0\s$ &$ 0.671$ &$0.0\s$ &$ 0.745$ &$0.0\s$ &$ 0.784$ &$0.0\s$ &
  $ 0.142$ &$0.0\s$\\
$(\widetilde{Q}_{22})_n$ &
$ 0.0\s$ &$0.730$ &$ 0.0\s$ &$0.706$ &$ 0.0\s$ &$0.693$ &$ 0.0\s$ &$0.716$ &
  $ 0.0\s$ &$0.726$\\
$(\widetilde{P}_+)_p$ &
$-0.325$ &$0.0\s$ &$-0.246$ &$0.0\s$ &$-0.225$ &$0.0\s$ &$-0.240$ &$0.0\s$ &
  $-0.009$ &$0.0\s$\\
$(\widetilde{P}_-)_p$ &
$-0.026$ &$0.0\s$ &$-0.024$ &$0.0\s$ &$-0.021$ &$0.0\s$ &$-0.019$ &$0.0\s$ &
  $-0.001$ &$0.0\s$\\
$(\widetilde{Q}_{20})_p$ &
$ 0.525$ &$0.0\s$ &$ 0.540$ &$0.0\s$ &$ 0.520$ &$0.0\s$ &$ 0.467$ &$0.0\s$ &
  $ 0.016$ &$ 0.0\s$\\
$(\widetilde{Q}_{22})_p$ &
$ 0.0\s$ &$0.684$ &$ 0.0\s$ &$0.708$ &$ 0.0\s$ &$0.720$ &$ 0.0\s$ &$0.698$ &
  $ 0.0\s$ &$0.688$\\
\hline
$\delta_B$
&0.058 &0.064 &0.010 &0.028 &0.011 &0.008 &0.132 &0.007 &0.137 &0.010\\
$\delta_1$
&0.410 &0.059 &0.406 &0.025 &0.140 &0.004 &0.176 &0.005 &0.115 &0.008
\end{tabular}
\label{PRPA_Dy}
\end{table}

\begin{table}[htb]
\caption{
Similar to Table~\ref{PRPA_Os} but for Os isotopes.
Note that the nuclei $^{188,190}$Os have a triaxial shape in the ground state.
}
\begin{tabular}{crrrrrrrrrr}
&\multicolumn{2}{c}{$^{182}$Os} & \multicolumn{2}{c}{$^{184}$Os} &
 \multicolumn{2}{c}{$^{186}$Os} & \multicolumn{1}{c}{$^{188}$Os} &
 \multicolumn{1}{c}{$^{190}$Os} & \multicolumn{2}{c}{$^{192}$Os} \\
&\multicolumn{1}{c}{$\beta$} & \multicolumn{1}{c}{$\gamma$} &
 \multicolumn{1}{c}{$\beta$} & \multicolumn{1}{c}{$\gamma$} &
 \multicolumn{1}{c}{$\beta$} & \multicolumn{1}{c}{$\gamma$} &
 & & \multicolumn{1}{c}{$\beta$} & \multicolumn{1}{c}{$\gamma$} \\
\hline
$\hbar\Omega$& \multicolumn{1}{c}{1.57}& \multicolumn{1}{c}{1.12}&
 \multicolumn{1}{c}{1.51}& \multicolumn{1}{c}{0.85}& \multicolumn{1}{c}{1.47}&
 \multicolumn{1}{c}{0.42}& \multicolumn{1}{c}{0.59}& \multicolumn{1}{c}{0.52}&
 \multicolumn{1}{c}{1.64}& \multicolumn{1}{c}{0.61}\\
$\hbar\bar{\Omega}$& \multicolumn{1}{c}{1.85}& \multicolumn{1}{c}{1.15}&
 \multicolumn{1}{c}{1.78}& \multicolumn{1}{c}{0.86}& \multicolumn{1}{c}{1.73}&
 \multicolumn{1}{c}{0.43}& \multicolumn{1}{c}{0.61}& \multicolumn{1}{c}{0.53}&
 \multicolumn{1}{c}{2.02}& \multicolumn{1}{c}{0.62}\\
\hline
$(\widetilde{P}_+)_n$ &
$-0.014$& $0.0\s$& $ 0.265$& $0.0\s$& $ 0.174$& $0.0\s$& $-0.014$& $-0.125$&
 $ 0.236$& $0.0\s$\\
$(\widetilde{P}_-)_n$ &
$ 0.001$& $0.0\s$& $-0.081$& $0.0\s$& $-0.162$& $0.0\s$& $ 0.208$& $ 0.113$&
 $-0.082$& $0.0\s$\\
$(\widetilde{Q}_{20})_n$ &
$ 0.009$& $0.0\s$& $-0.186$& $0.0\s$& $-0.278$& $0.0\s$& $ 0.178$& $ 0.504$&
 $ 0.561$& $0.0\s$\\
$(\widetilde{Q}_{22})_n$ &
$ 0.0\s$& $0.690$& $ 0.0\s$& $0.699$& $ 0.0\s$& $0.705$& $ 0.658$& $ 0.472$&
 $ 0.0\s$& $0.705$\\
$(\widetilde{P}_+)_p$ &
$ 0.999$& $0.0\s$& $ 0.905$& $0.0\s$& $ 0.847$& $0.0\s$& $-0.049$& $-0.047$&
 $ 0.381$& $0.0\s$\\
$(\widetilde{P}_-)_p$ &
$ 0.031$& $0.0\s$& $-0.069$& $0.0\s$& $-0.101$& $0.0\s$& $ 0.099$& $ 0.064$&
 $-0.117$& $0.0\s$\\
$(\widetilde{Q}_{20})_p$ &
$-0.028$& $0.0\s$& $-0.256$& $0.0\s$& $-0.372$& $0.0\s$& $ 0.184$& $ 0.509$&
 $ 0.681$& $0.0\s$\\
$(\widetilde{Q}_{22})_p$ &
$ 0.0\s$& $0.724$& $ 0.0\s$& $0.715$& $ 0.0\s$& $0.709$& $ 0.668$& $ 0.479$&
 $ 0.0\s$& $0.710$\\
\hline
$\delta_B$
 &0.072 &0.002 &0.067 &0.0005 &0.068 &0.00005 &0.001
   &0.001 &0.313 &0.0004\\
$\delta_1$
 &0.064 &0.001 &0.087 &0.0003 &0.156 &0.00004 &0.835
   &0.657 &0.329 &0.0003
\end{tabular}
\label{PRPA_Os}
\end{table}

\end{document}